\documentstyle[11pt]{article}

\input{psfig}

\def\text{}
\renewcommand{\baselinestretch}{1.75} 

\newcommand{\al}{\alpha}

\newcommand{\th}{\theta}

\newcommand{\la}{\lambda}

\newcommand{\De}{\Delta}

\newcommand{\La}{\Lambda}

\def\NSzero{ {\bigcirc \!\!\!\!\! 0}\,\,}
\def\NSone{ {\bigcirc \!\!\!\!\! 1}\,\,}
\def\NStwo{ {\bigcirc \!\!\!\!\! 2}\,\,}
\def\NSthree{ {\bigcirc \!\!\!\!\! 3}\,\,}
\def\NSfour{ {\bigcirc \!\!\!\!\! 4}\,\,}
\def\NSfive{ {\bigcirc \!\!\!\!\! 5}\,\,}
\newcommand{\be}{\begin{eqnarray}}
\newcommand{\ee}{\end{eqnarray}}

\newcommand{\tf}{\tilde{f}}

\newcommand{\pr}{\partial}

\newcommand{\np}{\newpage}
\newcommand{\hs}{\hspace}
\newcommand{\vs}{\vspace}
\newcommand{\nl}{\newline}
\newcommand{\nn}{\nonumber}

\def\tstwothirds{ {\textstyle {2\over 3}} }
\def\tshalf{ {\textstyle {1\over 2}} }
\newcommand{\RR}{{\rm I\kern-1.6pt {\rm R}}}

\newcommand{\ZZ}{{\rm Z}\kern-3.8pt {\rm Z} \kern2pt}

\begin{document}

\thispagestyle{empty}

\vs*{-25mm}
\begin{flushright}
BRX-TH-463\\[-.2in]
BOW-PH-115\\[-.2in]
\end{flushright}
\setcounter{footnote}{0}

\begin{center}
{\Large{\bf Tests of M-Theory from ${\cal N}=2$ 
Seiberg-Witten Theory}}
\renewcommand{\baselinestretch}{1}
\small
\normalsize
\vspace{.2in}
\footnote{
Based on lectures by H.J. Schnitzer at the 
{\it Advanced School of Supersymmetry in the \\
\phantom{aaa}  Theories of Fields,  Strings and Branes},
 University of 
Santiago de Compostela, Spain, July 1999.}\\

Isabel P. Ennes
\footnote{
Research supported 
by the DOE under grant DE--FG02--92ER40706.}\\
Martin Fisher School of Physics\\
Brandeis University, Waltham, MA 02254

\vspace{.1in}

Stephen G. Naculich
\footnote{
Research supported in part by the National Science Foundation under grant 
no.~PHY94-07194 through the \\
\phantom{aaa}  ITP Scholars Program.}\\
Department of Physics\\
Bowdoin College, Brunswick, ME 04011

\vspace{.1in}

Henric Rhedin\\
Celsius Consultants,
Chalmers Teknikpark\\
S-412 88 G\"oteborg, Sweden.

\vspace{.1in}

Howard J. Schnitzer\footnote{Research supported in part
by the DOE under grant DE--FG02--92ER40706.\\
{\tt \phantom{aaa} naculich@bowdoin.edu; Henric.Rhedin@celsius.se; 
ennes,schnitzer@brandeis.edu}\\}\\
Martin Fisher School of Physics\\
Brandeis University, Waltham, MA 02254

\vspace{.2in}

{\bf{Abstract}} 
\end{center}
\renewcommand{\baselinestretch}{1.75}
\small
\normalsize
\begin{quotation}
\baselineskip14pt
\noindent  

Methods are reviewed for computing the instanton expansion 
of the prepotential for ${\cal N}=2$ Seiberg-Witten (SW) theory
with {\it non}-hyperelliptic curves. These results, if compared
with the instanton expansion obtained from the microscopic Lagrangian,
will provide detailed tests of \hbox{M-theory}. 
We observe group-theoretic regularities of 
${\cal F}_{\rm 1-inst}$ which allow
us to ``reverse engineer" a SW curve for ${\rm SU}(N)$ gauge theory
with two hypermultiplets in the antisymmetric representation
and $N_f\leq 3$ hypermultiplets in the fundamental representations, 
a result not yet available by other methods. 
Consistency with \hbox{M-theory} requires a curve 
of infinite order, which we identify as a 
decompactified version of elliptic models of the type described 
by Donagi and Witten, Uranga, and others. This 
leads us to a brief discussion of some elliptic models 
that relate to our work. 
\end{quotation}

\np 

\setcounter{page}{1}

\noindent{\bf 1. ~Objectives}
\renewcommand{\theequation}{1.\arabic{equation}}
\setcounter{equation}{0}

It is our purpose to present a method for 
obtaining precise tests of \hbox{M-theory} using 
\hbox{${\cal N}=2$} Seiberg-Witten 
 supersymmetric (susy) gauge theory \cite{SeibergWitten}. Although 
the string community believes in \hbox{M-theory}, 
it must nevertheless be subjected to detailed
verification for the same reasons that one 
subjects quantum electrodynamics to such precision tests as
the measurement and computation of $g-2$. In our context, 
\hbox{M-theory} provides SW curves for low-energy 
effective \hbox{${\cal N}=2$} 
susy gauge theories, which in principle allows one to 
compute the instanton expansion of the 
prepotential of the theory. The results
of this calculation must be compared with calculations of 
the instanton contributions to the prepotential 
from the microscopic Lagrangian. It is this comparison which provides 
the tests we are concerned with. It should be noted 
that what we are considering is the ability of \hbox{M-theory} 
to make detailed non-perturbative predictions for field 
theory, as we consider the limit in which gravity has decoupled.  

M-theory provides SW curves for effective \hbox{${\cal N}=2$} 
susy gauge theories with
hypermultiplets in the both fundamental representation \cite{Mtheory, Witten}
and in higher representations \cite{LandsteinerLopezLowe}. 
Since the (hyperelliptic) curves from the former were initially obtained
from purely field-theoretic considerations \cite{Everybody}, 
we regard 
these as {\it post}-dictions of \hbox{M-theory}, 
though the agreement is gratifying.  In order to obtain 
genuine tests of \hbox{M-theory} we need to consider 
situations for which it is not known how to obtain SW curves 
from field-theoretic arguments alone. Examples of this
kind are, for example,  \hbox{${\cal N}=2$} ${\rm SU}(N)$  gauge theory
with a hypermultiplet in the symmetric or 
antisymmetric representation 
\cite{LandsteinerLopezLowe}. In such cases \hbox{M-theory} gives the 
only known predictions 
of the relevant SW curves, which 
happen to be {\it non}-hyperelliptic curves. If one can extract 
the instanton expansion for these 
examples, and compare these to results from a microscopic 
calculation, one will have genuine 
tests of \hbox{M-theory}. The problem
we faced is that there were no known methods to obtain the 
instanton expansion. Our solution to this
issue will be one of the main themes of this 
review\cite{oneanti}--\cite{twoanti}. 

One of the intriguing aspects of SW theory is the connection
to integrable models: 
elliptic models in particular 
\cite{DonagiWitten}--\cite{Gorsky}. 
\hbox{M-theory} provides 
one method of constructing the 
spectral curves of elliptic models.
Another approach to these problems is that of geometric engineering
\cite{geomengineering}, which we will not discuss here.  We will
see that our efforts in understanding SW theory with 
non-hyperelliptic curves leads us in a natural 
way to the consideration of \hbox{M-theory} and elliptic models. 
Some aspects of this connection will be 
the second main theme of this paper. 

A schematic chart of some of these connections is shown in Figure 1.

\begin{picture}(810,400)(10,30)

\put(200,360){\framebox(80,50){{\shortstack{M-theory \\ (or geometric \\
engineering)}}}}
\put(240,360){\vector(0,-1){50}}
\put(200,260){\framebox(80,50){{\shortstack{Instanton \\ expansion of \\
prepotential}}}}
\put(270,310){\makebox(80,50){{\shortstack{SW curve for \\ Coulomb branch of
\\ ${\cal N}=2$ susy Yang-Mills}}}}
\put(240,260){\vector(0,-1){120}}
\put(170,90){\framebox(140,50){{\shortstack{Strong coupling results. \\ Not
available from \\ microscopic Lagrangian}}}}
\put(260,180){\framebox(150,50){{\shortstack{Instanton calculations \\ from
microscopic \\ Lagrangian}}}}
\put(280,260){\vector(1,-1){30}}
\put(283,257){\vector(-1,1){3}}
\put(280,227){\makebox(80,50){{Test curve}}}
\put(80,180){\framebox(130,50){{\shortstack{Non-stringy methods. \\
Integrable \\ theories, etc.}}}}
\put(200,260){\vector(-1,-1){30}}
\put(197,257){\vector(1,1){3}}
\put(130,227){\makebox(80,50){{Curve}}}
\put(180,143){\vector(0,-1){3}}
\put(180,149){\line(0,-1){3}}
\put(180,155){\line(0,-1){3}}
\put(180,161){\line(0,-1){3}}
\put(180,167){\line(0,-1){3}}
\put(180,173){\line(0,-1){3}}
\put(180,179){\line(0,-1){3}}
\put(240,40){\makebox(0,0)[b]{\bf {Figure 1}}}

\end{picture}
\eject

\noindent{\bf 2. ~Topics considered}
\renewcommand{\theequation}{2.\arabic{equation}}
\setcounter{equation}{0}

\noindent 
1) {\it Brief introduction} to 

a) \hbox{${\cal N}=2$} Seiberg-Witten theory.

b) \hbox{M-theory} construction of SW curves (Riemann surfaces).

\noindent 
2) \,a) Instanton expansion for {\it non}-hyperelliptic curves.

b) Tests + predictions of \hbox{M-theory}. 

\noindent
3) Reverse engineer a curve for 
\hbox{${\cal N}=2$} ${\rm SU}(N)$ susy gauge theory, with two antisymmetric 
representations and $N_f \leq 3$ hypermultiplets. 

\noindent
{\it Require} consistency with \hbox{M-theory} 
which implies a curve of infinite order.

\noindent
4) Relation to:

a) Elliptic models. 

b) Integrable models. 

\noindent
5) Some unsolved problems and concluding remarks. 

\noindent{\bf 3. ~Seiberg-Witten Theory}
\renewcommand{\theequation}{3.\arabic{equation}}
\setcounter{equation}{0}

We will be concerned with \hbox{${\cal N}=2$} susy 
Yang-Mills theory in $D=4$ dimensions, with gauge 
group ${\cal G}$, together with hypermultiplets 
in some representation $R$. This theory can be described 
by a microscopic Lagrangian 
\be
{\cal L}_{\rm micro}=\frac{1}{4 g^2}\, F_{\mu \nu} ^a\, F^{\mu \nu a}\,+
\, {\theta \over 32 \pi^2}\,F_{\mu \nu} ^a\,\tilde F^{\mu \nu a}\,
+\, D_\mu \phi^{+} D^\mu \phi\,+\, {\rm tr} [\phi,\phi^{+}]^2\,\nn \\
+\, {\rm fermion}\,+\,{\rm hypermultiplet}\,\, {\rm terms}, 
\label{one}
\ee 
with $\mu, \nu =1$ to $4$ and $a=1$ to dim ${\cal G}$.
The field strength $F_{\mu \nu}$ and the scalar field $\phi$ belong 
to the adjoint representation, as they are the 
bosonic components of the \hbox{${\cal N}=2$} 
gauge multiplet. The vacuum is described by the condition 
\be
[ \phi,\phi^{+}]\,=\,0, 
\label{two}
\ee
which implies $\phi^a\,=\,{\rm constant}$. 
One may rotate  $\phi^a$  to the Cartan 
subalgebra, in which case 
\be
{\rm diag}\,(\phi) \,=\, (a_1, a_2, \ldots)\,,  
\,\,\,\,\,\,\,\,\,{\rm with} \,\,\,\,\sum_i a_i=0.
\label{three}
\ee
If all the $a_i$ are distinct, 
this generically breaks ${\cal G}$ to ${\rm U(1)}^{\it rank \,\,{\cal G}}$.

1) If only $\phi$ acquires a vacuum expectation value (vev), we define this 
as the Coulomb branch.

2) If only the scalar fields in the matter hypermultiplets have a vev, 
this is the Higgs branch. 

3) There are also mixed branches. 

\noindent We will focus on the Coulomb branch of these theories. 

The breakthrough of Seiberg and Witten 
\cite{SeibergWitten}
was their 
formulation of the exact solution 
of low-energy \hbox{${\cal N}=2$} susy gauge 
theories in terms of an effective (Wilsonian) action 
accurate to two derivatives of the fields. 
In $D=4$, the SW program is  described
in terms of 
 \be
{\cal L}_{eff}\,=\,\frac{1}{4\pi} {\rm Im}\left(\int {\rm d}^4\th
\frac{\pr {\cal F}(A)}{\pr A_i}\bar{A_i}+
\frac{1}{2}\int {\rm d}^2\th\frac{\pr^2 {\cal F}(A)}{\pr A_i\,\pr A_j}
W^{\al}_i\,W_{\al,j}\right)\,+\,{\rm higher~derivatives}, \nn \\
\label{four}
\ee 
where $A^i$ are ${\cal N}=1$ chiral 
superfields ($i=1 \,\,\,{\rm to}\,\, \,{\rm rank}\,{\cal G}$), 
${\cal F}(A)$ is the holomorphic prepotential, and $W^i$ is the
gauge field strength. In components the effective action is 
\be
{\cal L}_{eff}\,=\,{1\over 4}  
{\rm Im} (\tau _{ij})\, F_{\mu \nu}^i F^{\mu \nu j}\,+\,
{1\over 4} {\rm Re} (\tau _{ij})\, F_{\mu \nu}^i \tilde F^{\mu \nu \,j} \nn \\ 
+ \pr_{\mu} (a^+)^j  \pr^{\mu} (a_D)_j\,+\,{\rm fermions},
\label{five}  
\ee
where in (\ref{five}) we have only 
exhibited bosonic components of the \hbox{${\cal N}=2$} 
gauge superfield. We define the order parameters $a_i$ as in (\ref{three}), 
$(a_D)_j =  {\pr {\cal F}(a)\over \pr a_j}$ denote the dual
order parameters, and 
\be
\tau_{ij}= {\pr^2{\cal F}(a)\over \pr a_i \pr a_j},
\label{matrix}
\ee
is the coupling 
or period matrix. Note that 
${\rm Im} (\tau_{ij}) \geq 0$ for positive kinetic energies. 

The holomorphic prepotential can be expressed in 
terms of a perturbative piece and infinite series 
of instanton contributions as 
\be
{\cal F}(A)\,=\, {\cal F}_{\rm classical}(A)\,
+\,{\cal F}_{\rm 1-loop}(A)\,+\,\sum _{d=1}^{\infty} \Lambda^{(2N-I(R))d}
{\cal F}_{\rm d-inst}(A),
\label{six}
\ee
where we have specialized the 
instanton terms to ${\rm SU}(N)$, since we will concentrate on results 
for that group. Note that due 
to a non-renormalization theorem, the perturbative expansion for
(\ref{six}) terminates at \hbox{1-loop}, 
though there is an infinite series of non-perturbative instanton contributions. 
In (\ref{six}), $\Lambda$ is the 
quantum scale (Wilson cutoff) and $I(R)$ is the Dynkin
index of matter hypermultiplet(s) of representation $R$. Further 
\be
& &{\cal F}_{\rm 1-loop}(a)\,=\,\frac{i}{4 \pi}\sum_{\al \in \De_+}
(a\cdot \al)^2\,{\rm log}
\left(\frac{a\cdot \al}{\La}\right)^2\nn \\
& &-\frac{i}{8 \pi}\sum_{w \in W_G}\sum_{j=1}^{N_f}(a\cdot w+m_j)^2\,
{\rm log}\left(\frac{a\cdot w+m_j}{\La}\right)^2,
\label{seven}
\ee
where $\alpha$ ranges over 
the positive roots $\De_+$ of $\cal G$, $w$ runs over the weight
vectors for a hypermultiplet in the representation $R$, with mass $m$, and 
$a_i={\rm diag}(\phi)$ belongs to the 
Cartan subalgebra of ${\cal G}$. Notice that 
from perturbation theory 
$\tau_{ij}(a) \sim   {\rm log}(a_i-a_j)\,+\,...$ 
at large $a$, which is not single-valued. 

The Seiberg-Witten data which (in principle) 
allow one to reconstruct the prepotential
are:

\noindent
1) A suitable Riemann surface or algebraic curve, dependent on moduli 
$u_i$, or equivalently on the order parameters $a_i$.

\noindent
2) A preferred meromorphic 1-form $\lambda \equiv {\rm SW}$ differential.

\noindent
3) A canonical basis of homology cycles on the surface $(A_k, B_k)$.

\noindent
4) Computation of period integrals
\be
2\pi i a_k=\oint_{A_k}\la, \hs{15mm} 2\pi ia_{D,k}=\oint_{B_k}\la, 
\label{eight}
\ee
where recall 
$a_{D,k}\,=\,{{\pr {\cal F}{(a)}}\over {\pr a_k}}$ 
is the dual order parameter. The program 
is:

\noindent 
i) find the Riemann surface or 
algebraic curve appropriate to the given matter content,

\noindent 
ii) compute the period integrals, and

\noindent 
iii) integrate these to find ${\cal F}(a)$. 

What is known about the required Riemann surfaces?  For classical groups, 
with gauge multiplet and  $N_f$ 
hypermultiplets in the fundamental representation, where
$N_f$ is restricted by the requirement of asymptotic freedom, 
the curves encountered are {\it all} 
hyperelliptic. That is, they are all of the form \cite{Everybody}
\be
y^2\,+\,2 A(x)\,y\,+\,B(x)\,=\,0,
\label{nine}
\ee
where the coefficient functions depend 
on the moduli, or order parameters. It is important
to note that all curves in this class can be found from 
field-theoretic considerations, and do not require
\hbox{M-theory} for their derivation. 
On the other hand, for ${\rm SU}(N)$ with matter content 

\noindent
a) one antisymmetric representation 
and $N_f \leq N+2$ fundamental representations, or

\noindent
b) one symmetric representation and $N_f \leq N-2$ fundamental representations,

\noindent 
the appropriate curves are {\it not} hyperelliptic, but are cubic, of the form
\cite{LandsteinerLopezLowe}:
\be
y^3\,+\,2 A(x)\,y^2\,+\,B(x)\,y\,+\epsilon(x)\,=\,0.
\label{ten}
\ee
It should be emphasized that (\ref{ten}) has only been 
obtained by \hbox{M-theory}. The curve has
{\it not} been obtained by other methods. Therefore, 
any predictions of ${\cal L}_{eff}$, using 
(\ref{ten}), when compared with the analogous predictions of 
${\cal L}_{\rm micro}$
 should be considered genuine tests of \hbox{M-theory}. 
 The extraction of instanton predictions from
 curves of the form (\ref{ten}) will be 
 the concern of the first-half of these lectures. 
 
 In the second-half we will discuss 
 ${\rm SU}(N)$ gauge theory with two antisymmetric representations and 
 $N_f\leq 3$ for which {\it no} curve 
has yet been derived by {\it any} methods. We will
 ``reverse engineer" a curve, using 
observed group-theoretic regularities of ${\cal F}_{\rm 1-inst}$, and 
 demand consistency with \hbox{M-theory}. 
This will force us to consider a curve of infinite order. In so doing, 
 we will obtain a decompactified 
version of an elliptic  model of the type of 
 Donagi and Witten, Uranga,
 and others \cite{DonagiWitten}--\cite{MartinecWarner}. 
 This will lead to a brief discussion of elliptic models, and
 integrable models, as they relate to our work. 
 
 The main task in extracting instanton 
predictions from curves such as (\ref{ten}) 
 is the computation of the period 
 integrals (\ref{eight}), and the integration of  
 ${\pr {\cal F}(a)/ \pr a_k}$ to obtain ${\cal F}(a)$. There are two principal
 (complementary) methods to evaluate the period integrals 
 for {\it hyperelliptic} curves. These are: 
 
 \noindent
 1) Picard-Fuchs differential equations for the period integrals
 \cite{PicardFuchs}. This gives global 
 information throughout moduli space, but the complexity 
of the equations increases
 rapidly with rank  ${\cal G}$.
 
 \noindent
 2) Direct evaluation of the period integrals by asymptotic expansion
 \cite{DHokerKricheverPhong1, DHokerKricheverPhong2, DHokerPhong}. This method 
 is not limited by rank ${\cal G}$, and gives 
 results in ``natural" variables. One
 can only  easily obtain a few explicit terms of the 
 instanton expansion. However, there
 exists a nice recursion formula 
which can generate the instanton expansion recursively 
 from ${\cal F}_{\rm {1-inst}}$ 
 \cite{Chan}. (There are also other methods, involving 
 WDVV equations \cite{WDVV} or Whitham hierarchies
 \cite{Whitham}, which we do not consider here.)
 
 The problem we face is how to evaluate period integrals 
 \be
\oint \lambda\,=\,\oint {x dy\over y},
\label{eleven}
\ee
for non-hyperelliptic curves 
such as (\ref{ten}). For hyperelliptic curves, $y$ is given as a
square-root, found from the quadratic curve (\ref{nine}), 
and one can evaluate the resulting integral
(\ref{eleven}) by asymptotic expansion, for example
 \cite{DHokerKricheverPhong1, DHokerKricheverPhong2,DHokerPhong}. 
For the cubic curve, the exact solution is too 
complicated to be useable, while for curves of higher 
order, even exact solutions are not possible. 
Numerical solutions are of no interest, as we 
want to study the analytic behavior of ${\cal F}(a)$ 
on the order parameters.
In Sec. 5 we present a systematic method for extracting 
the instanton expansion 
for curves such as (\ref{ten}), 
which we argued is necessary if we are to test \hbox{M-theory} 
predictions for SW theory. First, 
we review in the next section how \hbox{M-theory} provides
Riemann surfaces for the SW problem.

\noindent{\bf 4. ~M-theory and the Riemann Surface}
\renewcommand{\theequation}{4.\arabic{equation}}
\setcounter{equation}{0}

The seminal work on this subject is by Witten \cite{Witten}, 
who considers IIA string theory 
lifted to \hbox{M-theory}. We summarize his discussion. It will be frequently 
convenient to use the language of IIA theory in describing the brane structure. 

For our first example consider 
${\rm SU}(N_1) \times {\rm SU}(N_2)$, described by IIA theory on $\RR^{10}$. 
The brane structure for this is 

\begin{picture}(430,200)(10,10)

\put(100,50){\line(0,1){150}}
\put(220,50){\line(0,1){150}}
\put(340,50){\line(0,1){150}}

\put(370,50){\vector(1,0){30}}
\put(404,48){$x_6$}

\put(100,40){\circle{12}}
\put(220,40){\circle{12}}
\put(340,40){\circle{12}}
\put(98,37){1}
\put(218,37){2}
\put(338,37){3}

\put(70,120){\vector(0,1){30}}
\put(69,153){$v$}
\put(42,110){$(x_4+ix_5)$}

\put(100,80){\line(1,0){9}}
\put(119,80){\line(1,0){9}}
\put(138,80){\line(1,0){9}}
\put(157,80){\line(1,0){9}}
\put(176,80){\line(1,0){9}}
\put(195,80){\line(1,0){9}}
\put(211,80){\line(1,0){9}}

\put(100,100){\line(1,0){9}}
\put(119,100){\line(1,0){9}}
\put(138,100){\line(1,0){9}}
\put(157,100){\line(1,0){9}}
\put(176,100){\line(1,0){9}}
\put(195,100){\line(1,0){9}}
\put(211,100){\line(1,0){9}}

\put(100,180){\line(1,0){9}}
\put(119,180){\line(1,0){9}}
\put(138,180){\line(1,0){9}}
\put(157,180){\line(1,0){9}}
\put(176,180){\line(1,0){9}}
\put(195,180){\line(1,0){9}}
\put(211,180){\line(1,0){9}}

\put(155,120){$\cdot$}
\put(155,125){$\cdot$}
\put(155,130){$\cdot$}
\put(155,115){$\cdot$}

\put(220,90){\line(1,0){9}}
\put(239,90){\line(1,0){9}}
\put(258,90){\line(1,0){9}}
\put(277,90){\line(1,0){9}}
\put(296,90){\line(1,0){9}}
\put(315,90){\line(1,0){9}}
\put(331,90){\line(1,0){9}}

\put(220,105){\line(1,0){9}}
\put(239,105){\line(1,0){9}}
\put(258,105){\line(1,0){9}}
\put(277,105){\line(1,0){9}}
\put(296,105){\line(1,0){9}}
\put(315,105){\line(1,0){9}}
\put(331,105){\line(1,0){9}}

\put(220,170){\line(1,0){9}}
\put(239,170){\line(1,0){9}}
\put(258,170){\line(1,0){9}}
\put(277,170){\line(1,0){9}}
\put(296,170){\line(1,0){9}}
\put(315,170){\line(1,0){9}}
\put(331,170){\line(1,0){9}}

\put(275,120){$\cdot$}
\put(275,125){$\cdot$}
\put(275,130){$\cdot$}
\put(275,135){$\cdot$}
\put(220,5){\makebox(0,0)[b]{\bf {Figure 2}}}

\end{picture}

The solid lines represent \hbox{NS 5-branes} and 
the dashed lines \hbox{D4-branes} suspended
between the \hbox {NS 5-branes}, with $N_1$ of these 
between ${\bigcirc \!\!\!\!\!1}\,$ and
${\bigcirc \!\!\!\!\!2}\,\,$, and $N_2$ between ${\bigcirc \!\!\!\!\!2}\,$ and 
${\bigcirc \!\!\!\!\!3}\,$.
The \hbox{NS 5-branes} are located at $x_7=x_8=x_9=0$, 
with world-volume $x_0$ to $x_5$. 
Classically, the \hbox{NS 5-branes} are at fixed 
values of $x_6$. The \hbox{D4-branes} have world-volume $x_0,
x_1, x_2, x_3, x_6$ with the ends of the \hbox{D4-branes} (classically) at fixed
values of $x_6$. Since the \hbox{D4-branes} are finite in extent in the $x_6$ direction, 
the macroscopic world-volume of the 
\hbox{D4-branes} is $(x_0, x_1, x_2, x_3)$, {\it i.e.} $d=4$. 
We consider the gauge theory on \hbox{D4-branes}. 

The  \hbox{NS 5-branes} are at definite positions in $x_6$ only classically. Intersection 
with the \hbox{D4-branes} creates a 
disturbance of the \hbox{NS 5-branes}, so that the $x_6$ position 
of a \hbox{NS 5-brane} should be measured at $v \sim \infty$, far from all disturbances. For large 
$v$, 
\be
\nabla^2 x_6(v, \bar v)\,=\,0, 
\label{twelve} 
\ee
which implies
\be
x_6\,=\,k {\rm log}\vert v \vert\,+\,{\rm const.}\,, 
\label{thirteen}
\ee
for a single \hbox{D4-brane}. 
For several \hbox{D4-branes} intersecting a given \hbox{NS 5-brane} 
from the left and right
\be
x_6\,=\,k\sum_{i=1}^{q_L} {\rm log}\vert v-a_i \vert\,-\,
k\sum_{i=1}^{q_R} {\rm log}\vert v-b_i \vert\,+\,{\rm const}.
\label{fourteen}
\ee
When there are $N$ parallel \hbox{D4-branes} suspended between a pair of \hbox{NS 5-branes}, 
generically
${\rm SU}(N)$ is broken to ${\rm U}(1)^{N-1}$. One can identify the logarithmic behavior in 
(\ref{fourteen}) with the large $v$ behavior of the $\alpha^{th}$ gauge theory 
${\rm SU}(N_{\alpha})$, with 
\be
{1\over g^2_{\alpha}(v)} \sim 
{x_6^{\alpha}(v)\,-\, x_6^{(\alpha-1)}(v) \over {\lambda}_{\rm IIA}} 
\sim {\rm log}\, v,
\label{fifteen}
\ee
where ${\lambda}_{\rm IIA}$ is the IIA string coupling.

The IIA string theory can be lifted to \hbox{M-theory}, 
as suggested in Figure 3: 

\begin{figure} [h]
\centerline{ 
\psfig{figure=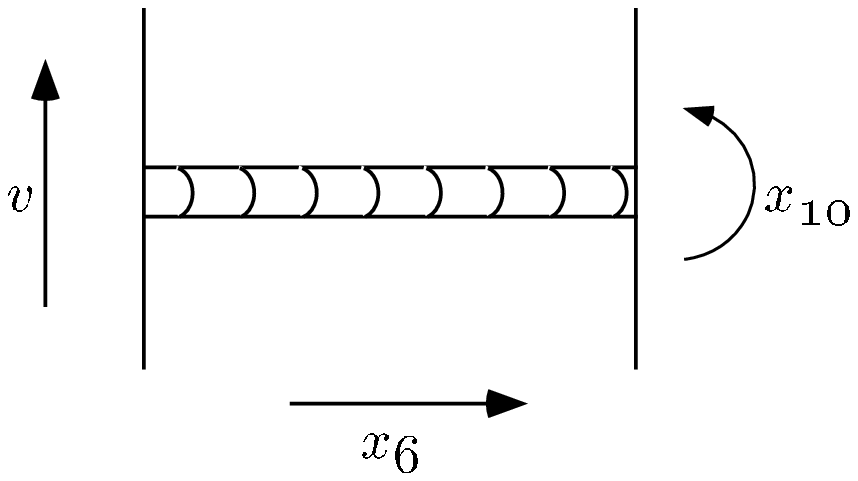,height=3cm,width=4cm}}
\begin{center}
{\footnotesize{\bf Figure 3}}
\end{center}
\end{figure}

Define 
\be
v&=&x_4+ix_5\,;\nn\\
s&=&{1\over R}\,(x_6+ix_{10}),
\label{sixteen}
\ee
where $R$ is the radius of the $11^{th}$ dimension ($x_{10}$), that is
\be
x_{10} \rightarrow x_{10}\,+2\pi R \qquad (\rm periodic).
\label{seventeen}
\ee
One then defines 
\be
\tau_{\alpha}(v)\,=\, {4\pi i \over g^2_{\alpha}}\,+\, {\theta_{\alpha}\over 2\pi}.
\label{eighteen}
\ee
Witten shows that
\be
-i\,\tau_{\alpha}(v) \sim (2k_{\alpha}\,-
\,k_{\alpha-1}\,-\,k_{\alpha+1})\, {\rm log}\,v 
\,=\, -(b_0)_{\alpha} \,\,{\rm log}\,v,
\label{nineteen}
\ee
where $-(b_0)_{\alpha}$ is the coefficient of 
the \hbox{1-loop} beta-function for the ${\alpha}^{th}$ gauge group. 

The brane picture lifted to \hbox{M-theory} is closely 
associated to the SW Riemann surface, since one can consider the
type IIA D4-brane in \hbox{M-theory} as an \hbox{M5-brane} 
wrapped on the $S^1$ of
(\ref{seventeen}). In fact, the type IIA setup with \hbox{D4-branes} and \hbox{NS 5-branes}
can be considered in \hbox{M-theory} as a {\it single} 
\hbox{M5-brane} with a very complicated
(6-dimensional) world-volume ${\RR}^{3,1} \times \Sigma$. One can then identify 
the brane picture for $\Sigma$ with the Riemann surface described by 
the SW curve. In this sense, the Riemann surface acquires a ``reality"
in \hbox{M-theory}, and is no longer just an auxiliary construct. 

Figure 2 describes an 
\hbox{${\cal N}=2$} ${\rm SU}(N_1)\times {\rm SU}(N_2)$ gauge theory
with matter hypermultiplets in the bifundamental representation 
$(N_1, \bar N_2)\,\oplus\,(\bar N_1, N_2)$. It is also
possible to add hypermultiplets in the fundamental representation.
To illustrate this, we consider \hbox{${\cal N}=2$}
SU$(N)$ gauge theory with $N_f <2N$.

\begin{picture}(430,200)(10,10)

\put(181,50){\line(0,1){150}}
\put(320,50){\line(0,1){150}}

\put(350,50){\vector(1,0){30}}
\put(384,48){$x_6$}

\put(110,120){\vector(0,1){30}}
\put(109,153){$v$}
\put(90,110){$(x_4+ix_5)$}

\put(181,80){\line(1,0){9}}
\put(200,80){\line(1,0){9}}
\put(219,80){\line(1,0){9}}
\put(238,80){\line(1,0){9}}
\put(257,80){\line(1,0){9}}
\put(276,80){\line(1,0){9}}
\put(295,80){\line(1,0){9}}
\put(311,80){\line(1,0){9}}

\put(181,100){\line(1,0){9}}
\put(200,100){\line(1,0){9}}
\put(219,100){\line(1,0){9}}
\put(238,100){\line(1,0){9}}
\put(257,100){\line(1,0){9}}
\put(276,100){\line(1,0){9}}
\put(295,100){\line(1,0){9}}
\put(311,100){\line(1,0){9}}

\put(181,160){\line(1,0){9}}
\put(200,160){\line(1,0){9}}
\put(219,160){\line(1,0){9}}
\put(238,160){\line(1,0){9}}
\put(257,160){\line(1,0){9}}
\put(276,160){\line(1,0){9}}
\put(295,160){\line(1,0){9}}
\put(311,160){\line(1,0){9}}

\put(245,120){$\cdot$}
\put(245,125){$\cdot$}
\put(245,130){$\cdot$}
\put(245,115){$\cdot$} 

\put(245,180){\framebox(5,5){$\cdot$}}
\put(205,60) {\framebox(5,5){$\cdot$}}
\put(245,10){\makebox(0,0)[b]{\bf {Figure 4}}}

\end{picture}

In Fig.~4, we have $N$ \hbox{D4-branes} 
suspended between two \hbox{NS 5-branes}. In addition, 
we have $N_f$ hypermultiplets in the fundamental 
representation denoted by ${\framebox(5,5){$\cdot$}}$ 
(in the figure, $N_f=2$), which describe \hbox{D6-branes} 
in $(x_0, x_1, x_2, x_3, x_7, x_8, x_9)$.
The bare masses of the hypermultiplets 
are given by the positions of the \hbox{D6-branes} 
in $v$. Define
\be
t\,=\,\exp(-s)\,=\,\exp\big[ -(x_6+ix_{10})/R\big].
\label{twenty} 
\ee
The curve which describes the positions of the \hbox{NS 5-branes} is
\be
t^2&=&{1\over 4} B(v)^2\,- c \prod_{j=1}^{N_f} (v-M_j),\nn\\
B(v)&=& d \prod_{j=1}^{N} (v-e_j),
\label{tone}
\ee
with $c$ and $d$ being constants. This curve is 
hyperelliptic. It agrees with the curve described earlier from
field theory \cite{Everybody}, 
so it is not an independent prediction of \hbox{M-theory}. 

Another ingredient that we will need is that of orientifold planes. In
particular, we will encounter $O6^{-}$ and $O6^+$ orientifold planes. An 
$O6^{\pm}$ orientifold is a \hbox{6-plane} that extends along the world-volume 
$(x_0, x_1, x_2, x_3, x_7, x_8, x_9)$ which produces a spacetime 
reflection. For
example, if the $O6^{\pm}$ is located at $x_4=x_5=x_6=0$, 
then these are fixed points
of spacetime under $(x_4,x_5,x_6) \rightarrow (-x_4,-x_5,-x_6)$. 
In other words, 
\be
(v,x_6) \rightarrow (-v,-x_6).
\label{ttwo}
\ee
(The orientifold also involves a world-sheet 
parity operation $\Omega$, and $(-1)^{F_L}$ which 
changes the sign of all left Ramond states. 
These considerations play no role in our discussion, 
but emphasize the perturbative nature of the orientifold). 
The $O6^-$ and $O6^+$ carry RR charge $-4$ and 
$+4$ respectively.

We have now all the elements to describe examples of \hbox{${\cal N}=2$} 
theories with non-hyperelliptic SW curves.  
Consider ${\rm SU}(N)$ gauge theory with either 
an antisymmetric or symmetric matter 
hypermultiplet \cite{LandsteinerLopezLowe}. The \hbox{M-theory} picture 
is

\begin{picture}(430,200)(10,10)

\put(100,50){\line(0,1){150}}
\put(220,50){\line(0,1){150}}
\put(340,50){\line(0,1){150}}

\put(370,50){\vector(1,0){30}}
\put(404,48){$x_6$}

\put(170,45){\vector(-1,0){30}}
\put(124,32){${\rm Mirror\ \ image}$}

\put(70,120){\vector(0,1){30}}
\put(69,153){$v$}

\put(216,124){$\otimes$}
\put(225,124){$O6$}

\put(100,80){\line(1,0){9}}
\put(119,80){\line(1,0){9}}
\put(138,80){\line(1,0){9}}
\put(157,80){\line(1,0){9}}
\put(176,80){\line(1,0){9}}
\put(195,80){\line(1,0){9}}
\put(211,80){\line(1,0){9}}

\put(100,100){\line(1,0){9}}
\put(119,100){\line(1,0){9}}
\put(138,100){\line(1,0){9}}
\put(157,100){\line(1,0){9}}
\put(176,100){\line(1,0){9}}
\put(195,100){\line(1,0){9}}
\put(211,100){\line(1,0){9}}

\put(100,161){\line(1,0){9}}
\put(119,161){\line(1,0){9}}
\put(138,161){\line(1,0){9}}
\put(157,161){\line(1,0){9}}
\put(176,161){\line(1,0){9}}
\put(195,161){\line(1,0){9}}
\put(211,161){\line(1,0){9}}

\put(155,120){$\cdot$}
\put(155,125){$\cdot$}
\put(155,130){$\cdot$}
\put(155,115){$\cdot$}

\put(220,90){\line(1,0){9}}
\put(239,90){\line(1,0){9}}
\put(258,90){\line(1,0){9}}
\put(277,90){\line(1,0){9}}
\put(296,90){\line(1,0){9}}
\put(315,90){\line(1,0){9}}
\put(331,90){\line(1,0){9}}

\put(220,152){\line(1,0){9}}
\put(239,152){\line(1,0){9}}
\put(258,152){\line(1,0){9}}
\put(277,152){\line(1,0){9}}
\put(296,152){\line(1,0){9}}
\put(315,152){\line(1,0){9}}
\put(331,152){\line(1,0){9}}

\put(220,170){\line(1,0){9}}
\put(239,170){\line(1,0){9}}
\put(258,170){\line(1,0){9}}
\put(277,170){\line(1,0){9}}
\put(296,170){\line(1,0){9}}
\put(315,170){\line(1,0){9}}
\put(331,170){\line(1,0){9}}

\put(275,120){$\cdot$}
\put(275,125){$\cdot$}
\put(275,130){$\cdot$}
\put(275,135){$\cdot$}
\put(220,10){\makebox(0,0)[b]{\bf {Figure 5}}}
\end{picture}

There are $3$ parallel \hbox{NS 5-branes} with $N$ \hbox{D4-branes} suspended 
between each. There is also an \hbox{$O6$-plane} 
on the central \hbox{NS 5-brane}, which therefore enforces the mirror 
symmetry (\ref{ttwo}) on the picture. In the 
absence of the orientifold, one would have 
${\rm SU}(N)\times {\rm SU}(N)$ with matter in the 
$(N, \bar N)\,\oplus\,(\bar N, N)$ representation. 
The orientifold ``identifies" the two ${\rm SU}(N)$ factors, 
projecting to the diagonal subgroup, giving one 
hypermultiplet in the antisymmetric representation for 
$O6^-$, or one hypermultiplet in the 
symmetric representation for $O6^+$. The curves for 
these situations have been worked out by 
Landsteiner and Lopez \cite{LandsteinerLopezLowe}. 
It is important to note that the orientifold 
induces a ${\ZZ}_2$ involution in the curve. 
The curves for these cases are shown in Figure~6.

\begin{picture}(500,275)(10,10)


\put(55,275){\line(1,0){380}}
\put(5,225){\line(1,0){380}}
\put(5,225){\line(1,1){50}}
\put(385,225){\line(1,1){50}}

\put(55,175){\line(1,0){380}}
\put(5,125){\line(1,0){380}}
\put(5,125){\line(1,1){50}}
\put(385,125){\line(1,1){50}}

\put(55,75){\line(1,0){380}}
\put(5,25){\line(1,0){380}}
\put(5,25){\line(1,1){50}}
\put(385,25){\line(1,1){50}}


\put(417,247){$y_1$}
\put(417,147){$y_2$}
\put(417,47){$y_3$} 

\put(212,37){$0$}

\put(80,37){$-e_{N-1}$}
\put(153,37){$-e_2$}
\put(180,37){$-e_1$}
\put(367,37){$-e_N$}

\put(55,255){$e_N$}
\put(237,255){$e_1$}
\put(262,255){$e_2$}
\put(337,255){$e_{N-1}$}


\put(340,150){\line(0,1){100}}
\put(265,150){\line(0,1){100}}
\put(240,150){\line(0,1){100}}
\put(60,150){\line(0,1){100}}

\put(215,50){\line(0,1){2}}
\put(215,58){\line(0,1){5}}
\put(215,68){\line(0,1){5}}
\put(215,78){\line(0,1){5}}
\put(215,88){\line(0,1){5}}
\put(215,98){\line(0,1){5}}
\put(215,108){\line(0,1){5}}
\put(215,118){\line(0,1){5}}
\put(215,128){\line(0,1){5}}
\put(215,138){\line(0,1){5}}
\put(215,148){\line(0,1){2}}

\put(90,50){\line(0,1){100}}
\put(165,50){\line(0,1){100}}
\put(190,50){\line(0,1){100}}
\put(375,50){\line(0,1){100}}


\put(110,47){$\cdot$}
\put(125,47){$\cdot$}
\put(140,47){$\cdot$}

\put(315,147){$\cdot$}
\put(300,147){$\cdot$}
\put(285,147){$\cdot$}

\put(110,147){$\cdot$}
\put(125,147){$\cdot$}
\put(140,147){$\cdot$}

\put(315,247){$\cdot$}
\put(300,247){$\cdot$}
\put(285,247){$\cdot$}

\end{picture}
{\footnotesize{\bf Figure 6} : The sheet 
structure for  the cubic curve corresponding to 
${\rm SU}(N)$ with either one symmetric or one antisymmetric representation.
For one symmetric representation, the curve is
\be
& & 
y^3 + f(x) y^2 + x^2 f(-x) L^2 y + x^6 L^6 =0,\nn\\
& & f(x) = \prod^N_{i=1} (x-e_i)\,,\,\,\,\,\,\,\,\,\,\, 
L^2\,=\,\Lambda^{N-2}\,,\nn\\
& &{\rm Involution:}\,\,\,y \rightarrow {{L^4 x^4} \over y}\,,
\,\,\,\,\,\,\,\,\,\, x \rightarrow -x. \nn
\ee
\noindent For one antisymmetric representation, the curve is
\be
& & y^3 + 2A (x) y^2 + B(x) y + L^6 =0,\nn\\
& & 2A(x) = \left[ f(x)  + 3  L^{2} \right]\,,\,\,\,\,\,\,\,\,\,\,\,\,
B(x)=L^{2} \left[ f(-x)  + 3L^{2} \right]\,,\nn\\
& & f(x) =x^2 \prod^N_{i=1} (x-e_i)\,,\,\,\,\,\,\,\,\,\,\,\,\,\,\,\,\,\,\,\,
L^2\,=\,\Lambda^{N+2}\,,\nn\\
& &{\rm Involution:} \,\,\,
y \rightarrow L^4/y\,,\,\,\,\,\,\,\,\,\,\,\, x \rightarrow -x. \nn
\ee
}

\eject

In Fig.~6 we have relabelled the variables 
$(t,v)\rightarrow(y,x)$ for convenience. 
Figure 6 can be visualized 
as type IIA or \hbox{M-theory} pictures by rotating so that $y$ is 
horizontal and $x$ is vertical. The Riemann surface
is a  three-fold branched covering of the Riemann sphere, 
with $N$ square-root \hbox{branch-cuts} connecting Riemann 
sheets $y_1$ and $y_2$, and also $y_2$ with $y_3$, with the 
central values of \hbox{branch-cuts} given by 
$e_1,\cdots, e_N$, with $\sum_{i=1}^{N}e_i=0$. Notice that the 
(classical) positions of the \hbox{D4-branes} correspond 
to the central values of the \hbox{branch-cuts}.
Finally, the curve is cubic, because one needs to describe 
the $y$ positions of the three \hbox{NS 5-branes}. 
The analytic problem at hand is how to compute the period 
integrals for such non-hyperelliptic curves. 
We describe our method for dealing with this in the next section. 

\noindent{\bf 5. ~Hyperelliptic Perturbation Theory}
\renewcommand{\theequation}{5.\arabic{equation}}
\setcounter{equation}{0}

We have developed a systematic scheme for 
the instanton expansion for prepotentials 
associated to non-hyperelliptic curves \cite{oneanti}--\cite{twoanti}. 
The method will be illustrated for the case of
${\rm SU}(N)$ gauge theory with one antisymmetric representation
\cite{oneanti} (see Fig.~6). The curve is
\be
y^3 + 2A (x) y^2 + B(x) y + \epsilon (x) =0,
\label{tthree}
\ee
where $L^2\,=\,\Lambda^{N+2}$, with $\Lambda$ the quantum scale of the theory, 
\be
\epsilon =L^6 \,\,;\,\,\,\,\,\,\,\,\,\,\,\,
2A(x)&=&\left[ f(x)  + 3\,L^{2} \right]\,,\nn\\
f(x) = x^2 \prod^N_{i=1} (x-e_i)\,\,;\,\,\,\,\,\,\,\,\,\,\,\,
B(x)&=&L^{2} \left[ f(-x)  + 3\,L^{2} \right].
\label{tfour}
\ee
It is fruitful to regard the last term 
$L^6$ in (\ref{tthree}) as a perturbation. The 
intuition  is that this involves 
a much higher power of the quantum scale in 
(\ref{tthree}) than the other terms, and 
geometrically it means separating the right-most 5-brane 
far from the remaining two \hbox{NS 5-branes}. Since 
the instanton expansion is an expansion in the 
quantum scale $\Lambda$, this idea has a chance of success.

To zeroth approximation we consider (\ref{tthree}) with $\epsilon=0$, 
which is then a hyperelliptic curve, 
and can be analyzed by previous available methods. This 
approximation gives ${\cal F}_{\rm 1-loop}$ 
correctly, but it is not adequate for ${\cal F}_{\rm 1-inst}$,
so one needs to go beyond the hyperelliptic approximation. Therefore, we
develop a systematic expansion in $\epsilon$, which we caution is not
the same as an expansion in $L$, as the coefficient functions $A(x)$ and $B(x)$
depend on $L$. To make the perturbation expansion look more symmetric define
\be
w = y + {2\over 3} A \; ,
\label{tfive}
\ee
so that (\ref{tthree}) is recast as 
\be
w^3 + \left(B - {\textstyle {4\over3}} A^2\right) w 
+ \left({\textstyle{16\over 27}} A^3 - {\tstwothirds} A B  + 
\epsilon \right) = 0.
\label{tsix}
\ee
The solutions to this equation satisfy 
\be
(w-w_1)(w-w_2)(w-w_3)
&=& 0,  \nonumber \\ [.1in]
w_1 + w_2 + w_3 
&=& 0,  \nonumber \\ [.1in]
w_1 w_2 + w_1 w_3 + w_2 w_3 
& = & B - {\textstyle {4\over3}} A^2,   \nonumber \\ [.1in]
w_1 w_2 w_3  
& = & -  {\textstyle{16\over 27}} A^3 + {\tstwothirds} A B  - \epsilon.
\label{tseven}
\ee
The zeroth approximation ($\epsilon=0$) is
\be
\bar w_1 & = & -{1\over 3} A - r, \nonumber \\
\bar w_2 & = & - {1\over 3}A + r, \nonumber \\
\bar w_3 & = & {2\over 3}A, \nn\\
r & = & \sqrt{A^2 - B}.
\label{teight}
\ee
The perturbation expansion for the solution to (\ref{tseven})  is 
\be
w_i =\bar w_i + \delta w_i &=& \bar w_i + 
\alpha_i \epsilon + \beta_i \epsilon^2 + \cdots\nn\\
\delta (w_1 + w_2 + w_3) & = & 0, \nonumber \\
\delta (w_1w_2 + w_2 w_3+ w_3w_1) & = & 0, \nonumber \\
\delta (w_1  w_2  w_3) & = & -\epsilon, 
\label{tnine}
\ee
where to first order
\be
\alpha_1 + \alpha_2 + \alpha_3 & = & 0, \nonumber \\
\bar w_1 \alpha_1 + \bar w_2 \alpha_2 + 
\bar w_3 \alpha_3  & = & 0, \nonumber\\
\bar w_2 \bar w_3 \alpha_1 + \bar w_1 \bar w_3 \alpha_2 
+ \bar w_1 \bar w_2 \alpha_3  & = & -1 \,,
\label{forty}
\ee
with solutions
\be
\alpha_1 & = & \frac{1}{(\bar w_1-\bar w_2)(\bar w_3-\bar w_1)}  
~=~ -  {1 \over 2r(A+r)} ~=~ - \, {A-r \over 2Br},
\nonumber \\[.1in]
\alpha_2& = & \frac{1}{(\bar w_2-\bar w_3)(\bar w_1-\bar w_2)}  
~=~ {1 \over 2r(A-r)} ~=~ {A+r \over 2Br},
\nonumber \\[.1in]
\alpha_3& = & \frac{1}{(\bar w_3-\bar w_1)(\bar w_2-\bar w_3)}  
=  - {1 \over B}\; . 
\label{fone}
\ee
One can go to the next order, where for example
\be
y_3  =  - \epsilon  {1 \over B} - 
\epsilon^2 \frac{2A}{B^3} + {\cal O} (\epsilon^3).
\label{ftwo}
\ee
It is to be observed in (\ref{teight}) and (\ref{fone}) 
that sheets $y_1$ and $y_2$ are
connected by square-root \hbox{branch-cuts}, 
as indicated by the factors of $r$, 
while (\ref{teight}), (\ref{fone}) and (\ref{ftwo}) show that sheet $y_3$ 
is decoupled from $y_1$
and $y_2$. In fact this is true 
in any finite order  of the $\epsilon$ perturbation
theory. However, we know that $y_2$ 
and $y_3$ are also connected by square-root \hbox{branch-cuts}
(see Fig.~6). This information is not really lost in our 
scheme. First note the expansion
\be
{1\over \sqrt{B^2 - A}}\,=\,{1\over B}\left[ 1+ {A\over {2 B^2}}+\cdots \right],
\label{fthree} 
\ee
is the structure observed in (\ref{ftwo}), which is the structure of the square 
root on sheet 3. 
More importantly, the involution
permutes the Riemann sheets, so that if $y_i(x)$ is a 
solution of (\ref{tthree}), then
\be
\bar y_i(x)\,=\,{L^4\over y_i(-x)},
\label{ffour}
\ee
is also a solution, with the property
\be
\bar y_1(x)&=&y_3(x), \nn \\
\bar y_2(x)&=&y_2(x), \nn \\
\bar y_3(x)&=&y_1(x),
\label{ffive}
\ee
so that (\ref{tnine})-(\ref{ftwo}), together with the involution captures
all the analytic structure of the solution. 

The SW differential appropriate to (\ref{tthree}) is 
\be
\lambda = x {dy \over y}.
\label{fsix}
\ee
Since sheet 3 is disconnected in any finite 
order of our perturbation expansion,
we need only consider $y_1$ and $y_2$. 
Let us label the SW differential for 
these two sheets $\lambda_1$ and $\lambda_2$ respectively, with
\be
\lambda_1 = x {dy_1 \over y_1},
\label{fseven}
\ee
and  $\lambda_2$ obtained from (\ref{fseven}) 
by $r \rightarrow {-r}$. The expansion
(\ref{tnine})  induces a comparable expansion for $\lambda$, whereby
\be
\lambda_1 = (\lambda_1)_I + (\lambda_1)_{II} + \cdots
\label{feight}
\ee
with
\be
\lambda_I
&=&  {x \left( {A^\prime\over A} - {B^\prime \over 2B} \right) \over 
\sqrt{1 - {B\over A^2}}}\,\, dx, \nn\\
\lambda_{II}
&=& -~ { L^6 \left(A - {B\over 2A}  \right)  \over
	 B^2 \sqrt{1 - {B\over A^2}} } \,\, dx, 
\label{fnine}
\ee
up to terms that do not contribute to period integrals. We note that 
$\lambda_I$ is the SW differential obtained 
from the hyperelliptic approximation
($\epsilon=0$) to 
(\ref{tthree}), and completely determines ${\cal F}_{\rm 1-loop}$, while
$\lambda_{II} \sim {\cal O}(L^2)$, so is of \hbox{1-instanton} order. Further
$(\lambda_1)_{III},\cdots$, contribute 
only to \hbox{2-instanton} order and higher, so
we may stop at (\ref{feight})-(\ref{fnine}) to \hbox{1-instanton} order.

In order to express the solutions to our problem 
with economical notation, we define 
certain ``{\it residue functions}", $R_k(x)$, 
$S(x)$, $S_0(x)$, and $S_k(x)$, where
\be
{R_k(x)\over (x-e_k)} \,=\, {3\over f(x)}\,
=\, \frac{3}{ x^2\prod_{i=1}^N (x-e_i) },
\label{fifty}
\ee
and
\be
S(x)\,=\,{4 f(-x)\over f^2(x)}\,=\,{S_0(x)\over x^2}\,=\,
{S_k(x)\over (x-e_k)^2}\,=\,{{4(-1)^N\prod^N_{i=1} (x+e_i)}\over  
{x^2\prod_{i=1}^N (x-e_i)^2}},
\label{fione}
\ee
so that 
\be
S_k(x)\,=\, {{4(-1)^N\prod^N_{i=1} (x+e_i)} \over 
{x^2\prod_{i\neq k} (x-e_i)^2}}.
\label{fitwo}
\ee
The functions $S(x)$ and $S_k(x)$ will play a crucial role for understanding 
of the general
features of the instanton expansion of SW problems. 

In order to calculate the period integrals, we must first 
locate the \hbox{branch-cuts} between
sheets $y_1$ and $y_2$. This is shown in Fig.~7

\bigskip

\begin{picture}(500,50)(10,10)

\put(50,40){\line(1,0){60}}
\put(180,40){\line(1,0){60}}
\put(310,40){\line(1,0){60}}

\put(130,38){$\cdot$}
\put(135,38){$\cdot$}
\put(140,38){$\cdot$}
\put(145,38){$\cdot$}

\put(260,38){$\cdot$}
\put(265,38){$\cdot$}
\put(270,38){$\cdot$}
\put(275,38){$\cdot$}

\put(40,30){$x^-_1$}
\put(100,30){$x^+_1$}
\put(74,30){$e_1$}
\put(76,40){\circle*{2}}

\put(170,30){$x^-_k$}
\put(230,30){$x^+_k$}
\put(204,30){$e_k$}
\put(206,40){\circle*{2}}

\put(300,30){$x^-_N$}
\put(360,30){$x^+_N$}
\put(334,30){$e_N$}
\put(336,40){\circle*{2}}
\put(220,0){\makebox(0,0)[b]{\bf {Figure 7}}}
\end{picture}
where the bare order parameters $\{e_i\}$ satisfy 
\be
\sum_i e_i=0
\label{fithree}
\ee
for a massless hypermultiplet.
Branch-cuts occur when $y_1=y_2$, so that 
\be
0 = A^2(x^\pm_k ) - B(x^\pm_k ) + 
\frac{L^6 A(x^\pm_k)}{2B(x^\pm_k)} + \ldots \; .
\label{fifour}
\ee
For small $L$, one may expand (\ref{fifour}) in powers of $L$, with the result
\be
x^\pm_k = e_k \pm L (S_k(e_k))^{1/2} + L^2 \left[{1\over 2} 
{{\pr S_k}\over {\pr x}}(e_k) -R_k(e_k)\right],
\label{fifive}
\ee
given in terms of (\ref{fifty})-(\ref{fitwo}).

In order to compute order parameters we need a 
canonical homology basis. For the order parameters $a_k$ we
have the basis $A_k$, as shown in Fig.~8.

\begin{figure} [h]
\centerline{
\psfig{figure=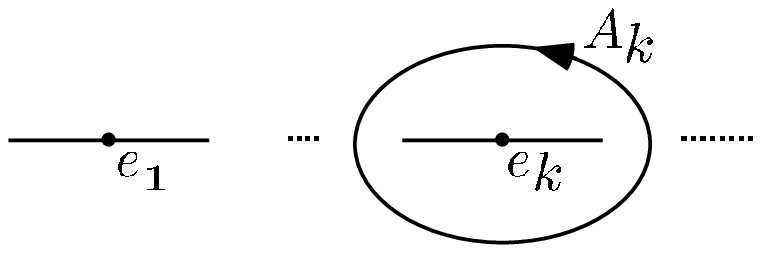,height=2cm,width=6.6cm}}
\begin{center}
{\footnotesize{\bf Figure 8}}
\end{center}
\end{figure}
We need the cycles $B_k$ for the dual order parameters
$a_{D,k}$ as shown in Fig.~9.

\begin{figure} [h]
\centerline{
\psfig{figure=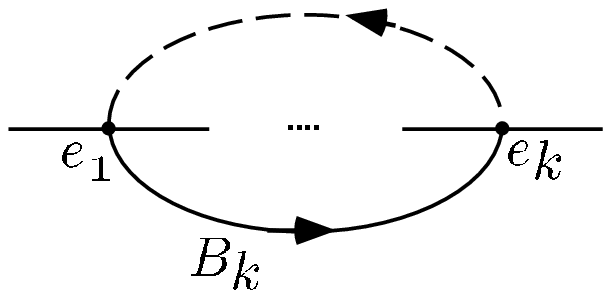,height=3.2cm,width=6.2cm}}
\begin{center}
{\footnotesize{\bf Figure 9}}
\end{center}
\end{figure}

The cycle $B_k$ connects sheets $y_1$ and $y_2$, with
the solid line on sheet $y_1$ and the dashed line on
$y_2$, with $B_k$ passing through the \hbox{branch-cut} 
as shown. 
To compute (\ref{eight}), one only needs 
$(\lambda_1 - \lambda_2)$, 
so that we only need keep terms odd 
under $r \rightarrow -r$, with $r$ as in (\ref{teight}). 

The order parameter is 
\be
2\pi{i} \, a_k & = & \oint_{A_k} \lambda  \nonumber \\
	& \simeq & \oint_{A_k} \left(\lambda_I 
+ \lambda_{II} \,+\cdots \right) \nonumber \\
& = &
\oint_{A_k} dx \left[
\frac{x \left( \frac{A^\prime}{A} - \frac{B^\prime}{2B} \right)}
{\sqrt{ 1-\, \frac{B}{A^2}}} 
 - L^6 { \left(A - {B\over 2A}  \right)  \over
	 B^2 \sqrt{1 - {B\over A^2}} }
\right].
\label{fisix}
\ee
The second term does not contribute to ${\cal O}(L^2)$, 
as there are no poles at $x=e_k$. Thus to our order
\be
a_k = e_k + L^2 \left[ \frac{1}{4} \frac{\pr S_k}{\pr x} (e_k)
- R_k (e_k) \right] + \cdots
\label{fieight}
\ee

The computation of the dual order parameter is considerably more complicated, 
with 
\be
2\pi{i} \, a_{D,k}\,=\,  \oint_{B_k} (\lambda_I + \lambda_{II}+\cdots).  
\label{finine}
\ee
The hyperelliptic approximation to (\ref{finine}) gives 
\be
2\pi{i} \, (a_{D,k})_{I}\, = \, 2 \int^{x^-_k}_{x^-_1} dx  
\left[
{{x \left( \frac{A^\prime}{A} - \frac{B^\prime}{2B} \right)}\over 
{\sqrt{ 1-\, \frac{B}{A^2}}}} \right]\,=\, \sum_{m=0}^{\infty} I_m,
\label{sixty}
\ee 
where the series is obtained by expanding the square-root in the denominator of 
(\ref{sixty}) with 
\be
I_m =  
\frac{2 \Gamma \left( m+ \frac{1}{2} \right) }
{\Gamma \left( \frac{1}{2} \right) \Gamma (m+1)} \;
\, \int^{x^-_k}_{x^-_1} dx \  x \,
\left( \frac{A^\prime}{A} -  \frac{B^\prime}{2B} \right)
\left( \frac{B}{A^2} \right)^m \; .
\label{sione}
\ee
Since we are only interested in computing $a_{D,k}$ accurate to 
${\cal O}(L^2)$ [\hbox{1-instanton} order], 
and since $B={\cal O}(L^2)$, one might 
naively expect that one need keep 
only $m=0$ and $m=1$ terms in (\ref{sixty})-(\ref{sione}).  
However, this is wrong, as the integrations
produce terms with $(1/L)^p$, for $p=1\,\,{\rm  to}\,\,\infty$. 
Therefore we need consider
all $m$ in (\ref{sixty}), and then sum  the series. 
The result of computing (\ref{sixty})-(\ref{sione})
is
\be
2\pi{i} \, (a_{D,k})_{I}&=& 
\oint_{B_k} \lambda_{I}\,=\, 
[N  + 2 + (N+2) \log (-1) + 2 \log L] \,a_k  
				\nonumber \\ [.1in]
&& - 2 \sum_{i\neq k}^N (a_k-a_i) \log (a_k-a_i) +
      \sum_{i=1}^N (a_k+a_i)  \log (a_k+a_i) -  2 a_k \log a_k 
				\nonumber \\ [.1in]
&& + L^2 \left[ - {1\over 4} \frac{\pr S_0}{\pr x} (0) \log a_k
	- {1\over 4} \sum_{j=1}^N  
	\frac{\pr S_j}{\pr x} (a_j) \log  \left( a_k+a_j \right)  \right.
\nonumber \\ [.1in]
&& + \left. {1\over 4} \frac{\pr S_k}{\pr x} (a_k) \;
+ {1\over 4} \frac{S_0(0)}{a_k}
- {1\over 2}  \sum_{i\neq k}^N \frac{S_i(a_i) }{a_k-a_i}\right] +{\cal O}(L^4).
\label{sitwo}
\ee
The terms of ${\cal O}(L^2)$ are \hbox{1-instanton} 
contributions to the dual order parameter. 
The presence of unallowed $L^2 \,{\rm log}\,a_k$ type terms indicates that 
(\ref{sitwo}) cannot be the complete contribution 
to $(a_{D,k})$ at \hbox{1-instanton} order.
Indeed, the second term in (\ref{finine}) is crucial 
for obtaining the correct result. This 
correction to the hyperelliptic approximation gives
\be
(2\pi i a_{D,k})_{II}
& = & \oint_{B_{k}} \lambda_{II}
			 \nonumber\\ [.1in]
& = & - 2L^2 \int^{x^-_k}_{x^-_1} dx \,{C(x) \over D^2(x) } 
		\nonumber\\ [.1in]
& =  & - L^2 \int^{x^-_k}_{x^-_1} dx \frac{\prod_i (x-e_i)}
{x^2\prod_i(x + e_i)^2}
         \nonumber\\ [.1in]
& = & {1\over 4} L^2   \left[
	 \frac{\pr S_0}{\pr x} (0) \log  a_k  
+ \sum^N_{j=1} \frac{\pr S_j}{\pr x} (a_j) \log  (a_k + a_j)
+  \frac{S_0(0)}{a_k} 
+ \sum^N_{j=1} \frac{S_j(a_j)}{a_k+a_j} 
\right] \;.\nn \\
\label{sithree}
\ee

Notice that the $L^2 \,{\rm log} a_k$ type terms in 
(\ref{sithree}) cancel those in (\ref{sitwo}).
The final result for $a_{D,k}$ to our order is given by  
(\ref{sitwo}) and (\ref{sithree}),
with 
\be
2 \pi i a_{D,k}
& = &
[N + 2 + (N+2) \log (-1) + 2 \log L]\, a_k  
				\nonumber \\ [.2 in]
&&-  2 \sum_{i\neq k} (a_k-a_i) \log (a_k-a_i) +
      \sum_i (a_k+a_i)  \log (a_k+a_i) -  2 a_k \log a_k 
				\nonumber \\ [.2 in]
&&+ L^2 \left[ {1\over 4} \frac{\pr S_k}{\pr x} (a_k) \;
   	+ {1\over 2}  \frac{S_0(0)}{a_k}
	- {1\over 2}\, \sum_{i\neq k} \frac{S_i(a_i) }{a_k-a_i} 
	+ {1\over 4}  \sum^N_{j=1} \frac{S_j(a_j)}{a_k+a_j} \right]
				\nonumber \\ [.2in]
&&  - ( k \to 1 ),
\label{sifour}
\ee
where the first and second rows of (\ref{sifour}) 
give the classical and \hbox{1-loop}
result respectively. It is possible to write (\ref{sifour}) as:
\be
2 \pi i a_{D,k} = 2 \pi i {\pr \over {\pr a_k}} \big[{\cal F}_{\rm classical} 
\,+\,{\cal F}_{\rm 1-loop}\big]\,
+\, L^2 \,{\pr \over {\pr a_k}} \big[-{1\over 2}S_0(0)\,+
\,{1\over 4}\sum_{k=1}^N S_k(a_k) \big],
\label{sifive}
\ee
in accord with the integrability required 
of the dual order parameter [cf (\ref{matrix})].
Thus we finally find for the prepotential for ${\rm SU}(N)$ gauge theory
with one massless antisymmetric hypermultiplet
\cite{oneanti}
\be
 {\cal F}_{\rm classical} + {\cal F}_{\rm 1-loop} 
&=& \frac{1}{4 \pi i} 
\left[ {{ 3 \over 2}} (N+2) 
+ (N+2)   {\rm log} (-1)  + 2 \,{\rm log}\,2 \right] \sum_j a_j^2
 \label{sisix} \\ [.2 in]
 && +  {i\over 8 \pi} 
\left[ \sum^N_{i,j=1} (a_i-a_j)^2 \log {(a_i-a_j)^2\over \Lambda^2}  
 - \sum_{i<j} (a_i + a_j)^2 \log {(a_i+a_j)^2\over \Lambda^2} \right],
 \nonumber
\ee
and
\be
{\cal F}_{\rm 1-inst} 
 = {1\over 2\pi i}
 \left[ - \frac{1}{2} \, S_0(0) +  \frac{1}{4} \, \sum_k S_k (a_k) \right].
\label{siseven}
\ee

Eq.~ (\ref{siseven}) is a prediction of \hbox{M-theory} which may be tested 
against microscopic calculations. 
This is presently possible for ${\rm SU}(N)$ with $N\leq 4$, since

\noindent
${\rm SU}(2)\, +\,{\rm antisymmetric} = {\rm SU}(2)$ (pure gauge theory);

\noindent
${\rm SU}(3)\, +\,{\rm antisymmetric} = {\rm SU}(3) \,+\,$ 1 fundamental;

\noindent
${\rm SU}(4)\, +\,{\rm antisymmetric} = {\rm SO}(6) \,+\, $1 fundamental.

In each of these three cases,  (\ref{siseven})  
agrees with \hbox{1-instanton} calculations 
from ${\cal L}_{\rm micro}$ \cite{instanton}. 
For $N \geq  5$, (\ref{siseven})
should be regarded as predictions of \hbox{M-theory}, awaiting testing. The
fact that  (\ref{siseven})
agrees with microscopic calculations, when available, after 
a long derivation, with distinct methods, is already impressive. 

There are further applications of hyperelliptic perturbation theory, where
the analysis is very similar to that 
sketched above. 
For ${\rm SU}(N)$ gauge theory with a hypermultiplet 
in the symmetric representation, which is also described by a cubic 
SW curve \cite{LandsteinerLopezLowe}, 
one obtains \cite{onesym}
\be
{\cal F}_{\rm 1-inst}\,=\, {1\over 8\pi i} \sum_{k=1}^N S_k(a_k),
\label{sieight}
\ee
where
\be
S_k(a_k)\,=\, 
{{4 (-1)^N a_k^2 \prod_{i=1}^N (a_k+a_i)}\over 
{\prod_{i\neq k}^N (a_k-a_i)^2 }}.
\label{sinine}
\ee
One may also add hypermultiplets 
in the fundamental representation. 
Moreover, 
one may consider hypermultiplets with non-zero masses. 
For ${\rm SU}(N)$ gauge theory with an antisymmetric representation
and $N_f< N+2$, which is described by a cubic
SW curve \cite{LandsteinerLopezLowe}, one finds \cite{nonhyper}:

\be
2\pi i {\cal F}_{\rm 1-inst}={1\over4}\sum_{k=1}^{N} S_k(a_k)-{1\over2}S_0(0),
\label{seventy}
\ee
where 
\be
S_k(a_k)={4(-1)^N\prod_{j=1}^{N_f}(a_k+M_j)\prod_{i=1}^{N}(a_k+a_i+m)\over 
(a_k+{1\over2}m)^2 \prod_{i\not=k}^N(a_k-a_i)^2},
\label{sevone}
\ee

\be
S_0(0)={4(-1)^N\prod_{j=1}^{N_f}(M_j-{1\over2}m)\over 
\prod_{k=1}^N(a_k+{1\over2}m)},
\label{sevtwo}
\ee
where $M_j$ ($m$) is the mass of the hypermultiplet 
in the fundamental (resp. antisymmetric) 
representation. Eqs. (\ref{sevone}) and (\ref{sevtwo}) 
agree with scaling limits taking 
$M_j$ and/or $m\rightarrow\infty$. 
Eqs.~(\ref{seventy})--(\ref{sevtwo}) provide additional tests 
of \hbox{M-theory}, since 

\noindent
${\rm SU}(2)\, +\,{\rm antisymmetric}\,+\, N_f ~{\rm fundamentals} = 
{\rm SU}(2)\,+N_f ~{\rm fundamentals}$,  

\noindent
${\rm SU}(3)\, +\,{\rm antisymmetric}\,+\, N_f ~{\rm fundamentals} 
= {\rm SU}(3) \,+\, (N_f+1) ~{\rm fundamentals}$, 

\noindent both of which agree with 
microscopic instanton calculations \cite{instanton}. However, 
for ${\rm SU}(N)\, +\,{\rm antisymmetric}\,+\, N_f ~{\rm fundamentals}$ with
$N\geq 4$, (\ref{seventy})--(\ref{sevtwo}) are predictions of \hbox{M-theory} 
which are as yet untested. 

The last example of an ${\rm SU}(N)$ theory with a cubic SW curve is 
${\rm SU}(N)\,+\,{\rm symmetric}\,+\,N_f~ {\rm fundamentals}$, 
with $N_f<N+2$. Here the result is
\cite{nonhyper}:
\be
2\pi i {\cal F}_{\rm 1-inst}={1\over4}\sum_{k=1}^{N} S_k(a_k),
\label{sevthree}
\ee
where 
\be
S_k(a_k)={4(-1)^N\,(a_k+{1\over2}m)^2\,
\prod_{j=1}^{N_f}(a_k+M_j)\,\prod_{i=1}^{N}(a_k+a_i+m)\over 
 \prod_{i\not=k}^N(a_k-a_i)^2}.
\label{sevfour}
\ee

Although (\ref{sevthree})--(\ref{sevfour}) have the correct 
scaling limits as $m$ or $M_j\rightarrow\infty$, 
these remain predictions of \hbox{M-theory} which 
have not been tested as yet. 

We have seen that whenever the predictions of the cubic SW curves 
obtained from \hbox{M-theory} have been tested, 
agreement has been found with those of microscopic calculations. 
However, there remain numerous further 
opportunities to subject \hbox{M-theory} predictions to testing. 

In the next section we discuss aspects of the ``universality" of the 
results of this section, and use 
those observations to construct a curve which is not yet obtainable 
from \hbox{M-theory}. 

\noindent{\bf 6. ~Universality}
\renewcommand{\theequation}{6.\arabic{equation}}
\setcounter{equation}{0}

In addition to the results for cubic SW curves discussed in 
the previous section, 
eqs. (\ref{siseven})-(\ref{sevfour}), we also wish to recall 
the results for a hyperelliptic curve, ${\rm SU}(N)\,
+\, N_f~{\rm fundamentals}$, with $N_f<2N$. That is
\cite{DHokerKricheverPhong1}, 
\be
8\pi i {\cal F}_{\rm 1-inst}=\sum_{k=1}^{N} S_k(a_k),
\label{sevfive}
\ee
where 
\be
S_k(a_k)={\prod_{j=1}^{N_f}(a_k+M_j)\over 
 \prod_{i\not=k}^N(a_k-a_i)^2}.
\label{sevsix}
\ee

If now one examines the cases treated in the previous section, 
together with (\ref{sevfive})--(\ref{sevsix}), 
one observes certain universal features: 

\noindent $(i)$ {The natural variables for this class of problems 
are the order parameters $\{a_k\}$ and 
not the 
gauge invariant moduli. 

\noindent $(ii)$ The \hbox{1-instanton} 
contribution to the prepotential can be written as 
\cite{onesym, nonhyper, DHokerKricheverPhong1}
\be
8\pi i {\cal F}_{\rm 1-inst}=\sum_{k=1}^{N} S_k(a_k),
\label{sevseven}
\ee
for ${\rm SU}(N)\,\,+\,\,N_f$ fundamentals or ${\rm SU}(N)$ 
$+$ symmetric $+$ $N_f$ fundamentals, and
\be
8\pi i {\cal F}_{\rm 1-inst}=\sum_{k=1}^{N} S_k(a_k)-2S_m(-m),
\label{seveight}
\ee
for ${\rm SU}(N)$ $+$ antisymmetric $+$ $N_f$ fundamental hypermultiplets
\cite{oneanti,nonhyper},
where the second term in (\ref{seveight}) 
removes a spurious singularity in ${\cal F}_{\rm 1-inst}$ as 
$a_k\rightarrow -m$. The generalization of 
(\ref{seveight}) will be important when we construct 
${\cal F}_{\rm 1-inst}$ for ${\rm SU}(N)$ gauge theory
with two antisymmetric hypermultiplets. 

We define (a posteriori) $S(x)$ which generalizes (\ref{fione}) as 
\be
S(x)\,=\,{S_k(x)\over (x-a_k)^2}
\,=\, {S_m(x)\over (x+m)^2}.
\label{sevnine}
\ee
Using the results of Sec. $5$, 
and of (\ref{sevsix}), we tabulate the resulting $S(x)$ in the first 
three entries of Table $1$. 
(We will discuss the $4^{th}$ row of Table $1$ shortly). 
It should be noted that 
Table $1$ includes all generic 
cases of asymptotically free \hbox{${\cal N}=2$} ${\rm SU}(N)$ gauge theories. 

A careful examination of the first three rows of 
Table $1$ leads to the following empirical rules for $S(x)$. 
$S(x)$ is given as the product of the 
following factors, each corresponding to 
a different \hbox{${\cal N}=2$} multiplet 
in a given representation of ${\rm SU}(N)$:

\noindent (1) Pure gauge multiplet factor 
\be
{1\over \prod_{i=1}^{N}(x-a_i)^2}.
\label{eighty}
\ee

\noindent (2) Fundamental representation. A factor 
\be
(x+M_j)
\label{eione}
\ee
for each hypermultiplet of mass $M_j$ in the fundamental representation.

\noindent (3) Symmetric representation. A factor 
\be
(-1)^N\,(x+m)^2\,\prod_{i=1}^{N}(x+a_i+2m)
\label{eitwo}
\ee
for each hypermultiplet of mass $2m$ in the symmetric representation.

\noindent (4) Antisymmetric representation. A factor 
\be
{(-1)^N\over(x+m)^2}\,\prod_{i=1}^{N}(x+a_i+2m)
\label{eithree}
\ee
for each hypermultiplet of mass $2m$ in the antisymmetric representation.

{}From these empirical rules, we predict $S(x)$ for ${\rm SU}(N)\,
+\,2~{\rm antisymmetric}\,+\, N_f~
{\rm fundamentals}$, with $N_f\leq 4$, as shown in the last entry of Table 1.

\begin{center}
\begin{tabular}{||c|c||}
\hline\hline
Hypermultiplet Representations &  $S(x)$\\
\hline\hline
        &{}                                                             \\
$N_f$ fundamentals
        & ${4\prod_{j=1}^{N_f}(x+M_j)\over \prod_{i=1}^N (x-a_i)^2}$    \\
(ref. \cite{DHokerKricheverPhong1})
        &{}                                                             \\
\hline
1\,\,{\rm symmetric}
        &{}                                                             \\
$+ N_f \,\,{\rm fundamentals}$
        & $ {4(-1)^N(x+m)^2
             \prod_{i=1}^N (x+a_i+2m) \prod_{j=1}^{N_f}(x+M_j)
                \over \prod_{i=1}^N (x-a_i)^2}$         \\
(ref. \cite{onesym, nonhyper})
        &{}                                                             \\
\hline
1\,\,{\rm antisymmetric}
        &{}                                                             \\
$+N_f\,\,{\rm fundamentals}$
        & ${4(-1)^N
        \prod_{i=1}^N (x+a_i+2m) \prod_{j=1}^{N_f}(x+M_j)
        \over (x+m)^2 \prod_{i=1}^N (x-a_i)^2}$                 \\
(ref. \cite{oneanti, nonhyper})
        &                                                                \\
\hline
2 \,\,{\rm antisymmetric}
        &{}                                                             \\
$ +N_f \,\,{\rm fundamentals}$
        & ${4\prod_{i=1}^{N}(x+a_i+2m_1) \prod_{i=1}^N(x +a_i+2m_2)
                \prod_{j=1}^{N_f}( x +M_j) \over ( x +m_1)^2 ( x +m_2)^2
                 \prod_{i=1}^N ( x -a_i)^2}$                            \\
(ref. \cite{twoanti})
        &{}                                                             \\
\hline

\end{tabular}
\end{center}
\label{tableone}
{\footnotesize{\bf Table 1}: The function $S(x)$
for SU$(N)$ gauge theory, with different matter content.
The hypermultiplets in the fundamental representation have masses $M_j$.
The symmetric or antisymmetric representation has mass $2m$.
If there are two antisymmetric representations,
their masses are $2m_1$ and $2m_2$}.
\vspace{1.5cm}

Given this $S(x)$, we then predict \cite{twoanti}
\be
8\pi i {\cal F}_{\rm 1-inst}= 
\sum_{k=1}^N S_k(a_k) -2 S_{m_1}(-m_1)-2S_{m_2}(-m_2), 
\label{eifour}
\ee
where $S_k(a_k)$ and $S_m$ 
are constructed from (\ref{sevnine}) and the $4^{th}$ entry of Table $1$. 
The last two terms of (\ref{eifour}) 
remove the spurious singularities as $a_k\rightarrow -m_1$ and
$a_k\rightarrow -m_2$, which 
generalizes (\ref{seveight}). 
Thus, from the observed regularities, we are able to predict ${\cal F}_
{\rm 1-inst}$ for ${\rm SU}(N)\,
+\,2~{\rm antisymmetric}\,+\,N_f~{\rm fundamentals}$, 
with $N_f\leq 3$, even though 
no SW curve is available from \hbox{M-theory}! 

The predictions of Table 1 and (\ref{eifour}) can be tested as follows:

\noindent
1)  ${\rm SU}(2)\, +\,2 \,{\rm antisymmetric} \,+ (N_f \leq 3)\, = 
{\rm SU}(2)\,+\,(N_f \leq 3)$.

\noindent
2)  ${\rm SU}(3)\, +\,2 \,{\rm antisymmetric} + (N_f \leq 3)\, = 
{\rm SU}(3) \,+\, (N_f\leq 5)$. 

\noindent
3) Limit $m_1$ or $m_2 \, \rightarrow \infty$ 
reduces to ${\rm SU}(N) + {\rm antisymmetric}+ (N_f \leq 3)$.  

In each of these 3 cases, our predicted ${\cal F}_{\rm 1-inst}$ 
finds agreement. 

The program we describe in the next section involves:

\noindent
1)  Assume that ${\cal F}_{\rm 1-inst}$ from Table 1 
and equation (\ref{eifour}) are correct. 

\noindent
2) Find a SW curve which gives 
\be
{\cal F}\,=\,  {\cal F}_{\rm classical }\,
+\, {\cal F}_{\rm 1-loop}\,+\, {\cal F}_{\rm 1-inst}. 
\nn
\ee

\noindent
3) Impose consistency with \hbox{M-theory}. 

It should be emphasized that there is no known derivation of 
the empirical rules of (\ref{eighty})-(\ref{eithree}). This is a problem that
deserves consideration from first principles. 

\noindent{\bf 7. ~Reverse Engineering a Curve}
\renewcommand{\theequation}{7.\arabic{equation}}
\setcounter{equation}{0}

Although there is no known SW curve for ${\rm SU}(N)$ gauge theory
with two antisymmetric
representations and $N_f\leq 3$ hypermultiplets, one can attempt to reverse
engineer a curve from the information in 
Table 1 and (\ref{eifour}). The strategy 
for the construction is 

\noindent
1) $ {\cal F}_{\rm classical}\,+\, {\cal F}_{\rm 1-loop}$ 
from perturbation theory. 

\noindent
2)  ${\cal F}_{\rm 1-inst}$ as predicted in Table 1 and  (\ref{eifour}).

\noindent
3) These two steps imply that 
$a_{D,k}= {{\pr {\cal F}} \over {\pr a_k}}$ 
is known to \hbox{1-instanton} accuracy. 

\noindent
4) Reproduce this expression from period integrals of a Riemann surface, to be
constructed from the above data.

\noindent
5) Ensure that the proposed Riemann surface is consistent with \hbox{M-theory}. 

To begin with  we postulate a quartic curve 
$$y^4\,+\,\cdots\,=\,0$$
Why? Since a cubic curve is needed for ${\rm SU}(N)$ gauge theory with 
a hypermultiplet in the symmetric or antisymmetric 
representations \cite{LandsteinerLopezLowe}, at least
a cubic curve is required. Further, Witten has shown \cite{Witten}
that for ${\rm SU}(N)\times {\rm SU}(N)\times
\buildrel{m~{\rm factors}}\over{\cdots}\times \,
{\rm SU}(N)$ the corresponding curve is 
$$y^{m+1}+\cdots =0,$$
which results from $m+1$ parallel \hbox{NS 5-branes}, and $N$ \hbox{D4-branes} 
suspended between neighboring pairs of 
\hbox{NS 5-branes}. However, for $({\rm SU}(N))^m$, with $m\geq 3$, we 
have shown \cite{product} that to attain \hbox{1-instanton} accuracy, 
one {\sl only} needs a quartic approximation 
$$ y^4+\cdots=0,$$
to the full $y^{m+1}$ curve. Therefore, we only need a quartic curve 
if we are trying to reproduce the 
prepotential to \hbox{1-instanton} accuracy. One can also show that the most 
general quartic curve consistent with \hbox{M-theory} is of the form
\cite{Witten, twoanti}
\be
& & L^4\,j_1(x)\,P_2(x)\,t^2\,+\,L\,P_1(x)\,t\,+
\,P_0(x)\,+\,L\,j_0(x)\,P_{-1}(x)\,{1\over t}\nn\\
&  & +\,
L^4\,j_0{}^2(x)\,j_{-1}(x)\,\,P_{-2}(x)\,{1\over t^2}=0,
\label{eifive}
\ee
where $j_n(x)$ are associated to the $N_f$ flavors in the fundamental 
representation, and $P_n(x)$ are 
associated to the positions of \hbox{D4-branes}. For our problem, 
\be
L^2=\Lambda^{4-N_f}.
\label{eisix}
\ee

Let us regard (\ref{eifive}) as the quartic approximation to a curve of 
infinite order. (We will justify 
this shortly when we demand that the brane picture for (\ref{eifive})
be consistent with \hbox{M-theory}, and 
an infinite number of reflections). Thus, (\ref{eifive}) is a truncation of 
\be
\sum_{n=1}^\infty L^{n^2}\,
\prod_{s=1}^{n-1}\,j_s{}^{n-s}(x)\,P_n(x)\,t^n\,+\,P_0(x)\,+\,
\sum_{n=1}^\infty L^{n^2}\,j_0{}^n(x)\,
\prod_{s=1}^{n-1}\,j_{-s}{}^{n-s}(x)\,P_{-n}(x)\,t^{-n}=0. \nn \\
\label{eiseven}
\ee

There should  be a symmetry between the two antisymmetric representations, 
which must appear in the curve. 
This is manifest by the symmetries
\be
P_n(x;m_1,m_2)&=&P_{-n}(x;m_2,m_1),\nn\\
j_n(x;m_1,m_2)&=&j_{-n}(x;m_2,m_1).\nn\\
\label{eieight}
\ee
The curve is then invariant under the involution 
\be
t\longrightarrow {j_0(x)\over t}; \qquad m_2 \longleftrightarrow m_1. 
\label{einine}
\ee

We wish to begin with the hyperelliptic perturbation 
expansion solutions to (\ref{eifive}), which is 
facilitated  by the  change of variables 
\be
t={1\over L P_1(x)}.
\label{ninety}
\ee
The curve becomes 
\be
{L^2\,j_1\,P_2\over P_1{}^2}\, y^4+ y^3+ P_0\,y^2+ 
L^2 j_0P_1P_{-1}y+L^6j_0^2j_{-1}P_1^2P_{-2}\,=\,0, 
\label{nione}
\ee
which is of the form
\be
\epsilon_{1}(x) y^4\,+\,y^3\,+2A(x)y^2\,+\,B(x)y\,+\,\epsilon_2(x)\,=\,0.
\label{nitwo}
\ee
To first order in $\epsilon_{1}$ and $\epsilon_{2}$, the solution 
for $y_1$ (sheet 1) is 
\be
y_1&=&-(A+r)\,-{(A+r)^3 \over 2r}\epsilon_1\,-\, 
{1\over {2r (A+r)}}\epsilon_2\,+\cdots\nn\\
&=& (y_1)_{I}\,+\,(y_1)_{II}\,+\cdots,
\label{nithree}
\ee
where 
\be
y_{I}\,=\,-(A+r)\,\,\,,\,\,\,\,\,\,\,\,\,\,r=\sqrt{A^2-B}.
\label{nifour}
\ee
Similarly, the SW differential is 
\be
\lambda\,=\, {x dy \over y}\ =\, \lambda_{I}\,+\,\lambda_{II}\,+\cdots
\label{nifive}
\ee
The problem is to find $P_n(x)$ and $j_n(x)$. 

{}From the period integral 
\be
\oint_{A_k} \lambda_{I} \,=\,\oint_{A_k} dx 
{{({A' \over A} -{B' \over {2B}})}\over 
\sqrt {1-{B\over A^2}}}, 
\label{nisix}
\ee
one has 
\be
a_k\,=\,e_k\,+{\cal O}(L^2),
\label{niseven}
\ee
for the order parameters. We {\it know} the dual order parameters,
to \hbox{1-instanton} accuracy, from Table 1 and (\ref{eifour}). 
They are related to the unknown 
coefficients as follows from the hyperelliptic approximation: 
\be
2\pi i (a_{D,k})_{I}&=&\oint_{B_k}\lambda_{I}  \label{nieight} \\
&=& 2L^2 \int_{x_1^-}^{x_k^-} dx 
\left[{{j_0(x) P_{-1}(x) P_1(x)}\over P_0^2(x)}\,
+\cdots\right] \label{ninine}\\
&=& -{1\over 2} L^2 \sum_{i\neq k}^{N} {S_i(a_i)\over (a_k-a_i)}\,+
\,\cdots \label{hundred} \\
&=& {L^2\over 2} \int_{x_1^-}^{x_k^-} dx \left[ 
\sum_{i=1}^N {S_i(x) \over (x-e_i)^2 }\,+\,\cdots\right].
\label{hone}
\ee
Eq. (\ref{ninine}) comes from the postulated 
form of the curve (\ref{nione}) and 
(\ref{nitwo}), while (\ref{hundred}) is obtained 
from Table 1 and  (\ref{eifour}), 
and (\ref{hone}) expresses (\ref{hundred})
as a period integral. There are other pieces of 
(\ref{nieight})-(\ref{hone})  not shown. 
We emphasize those features which lead to a solution for $P_n$ and $j_n$. 
Comparing (\ref{ninine}) with (\ref{hone}) leads to the postulate 
\be
{{j_0(x) P_{-1}(x) P_1(x)}\over P_0^2(x)}\,=
\, {1\over 4} S(x) \,+\,{\cal O}(L^2),
\label{htwo}
\ee
where the ${\cal O}(L^2)$ is what we call {\it subleading} terms. 

The correction to the hyperelliptic approximation gives
\be
2\pi i (a_{D,k})_{II}&=& 2 \int_{x_1^-}^{x_k^-} \lambda _{II}\nn\\
&=& -L^2 \int_{x_1^-}^{x_k^-} dx \left[{{j_1 P_0 P_2} \over P_{1}^2 }+ 
{{j_{-1} P_0 P_{-2}} \over P_{-1}^2 }  \right].
\label{hthree}
\ee 
Note that the same combination of coefficients appears 
in (\ref{htwo}) and (\ref{hthree}), but with the subscripts shifted 
as $n \rightarrow n+1$ or $n-1$.  The choice 
\be
{{j_1 P_0 P_2} \over P_1^2 }&=& 
{1\over 4}\,S(-x-2m_2)\,+\,{\cal O}(L^2), \nn\\
{{j_{-1} P_0 P_{-2}} \over P_{-1}^2 }&=& 
{1\over 4}\,S(-x-2m_1)\,+\,{\cal O}(L^2),
\label{hfour}
\ee
reproduces  a number of the terms of $(a_{D,k})$, with 
those remaining terms not obtained from 
(\ref{nieight})-(\ref{hone}) or (\ref{hthree})-(\ref{hfour})
assumed to come from subleading terms. 
Comparing (\ref{hfour}) with (\ref{htwo})
we argue that
\be
{{j_n P_{n-1} P_{n+1}} \over P_n^2 }&=& {1\over 4}\,
S({\rm known~reflection~and~shift~in}\,\, x)\,+\,{\cal O}(L^2). 
\label{hfive}
\ee 
This will have important geometrical consequences 
for the \hbox{M-theory} picture!

Remarkably, (\ref{htwo}), (\ref{hfour}) and (\ref{hfive}) 
can be uniquely solved for the 
{\it leading} coefficient functions,  with the first few
\be
P_0(x)&=& \prod_{i=1}^N (x-a_i) \,+\, {\cal O}(L^2), \nn\\
P_1(x)&=& (-1)^N (x+m_2)^{-2}\prod_{i=1}^N (x+a_i+2m_2) \,+
\, {\cal O}(L^2), \nn\\
P_2(x)&=&  (x+m_2)^{-6} (x+2m_2-m_1)^{-2}\prod_{i=1}^N (x-a_i+2m_2-2m_1) \,
+\, {\cal O}(L^2), 
\label{hsix}
\ee 
 and
 \be
 j_0(x)&=& \prod_{j=1}^{N_f}(x+M_j),\nn\\
 j_1(x)&=& (-1)^{N_f}\prod_{j=1}^{N_f}(x+2m_2-M_j),
\label{hseven}
 \ee
 together with (\ref{eieight}).
We shall associate the numerators of the $P_n(x)$ with the positions 
of the \hbox{D4-branes}, the denominators 
to that of $O6^-$ orientifold planes,
and $j_n(x)$ to the positions of the \hbox{D6-branes}. 

We now argue that  (\ref{eifive}) is incomplete if consistency with
\hbox{M-theory} is demanded, with the result of a curve of infinite order,
and therefore an infinite chain of \hbox{NS 5-branes} and orientifolds. 
To see the origin of this assertion, recall that the brane picture
for ${\rm SU}(N)$ gauge theory with an ${\rm antisymmetric \,representation}$ 
of mass $m$ is shown in Fig.~10, 
showing only the \hbox{NS 5-branes} and the $O6^-$
plane for clarity.

\begin{picture}(400,150)(10,10)

\put(100,20){\line(0,1){100}}
\put(200,20){\line(0,1){100}}
\put(300,20){\line(0,1){100}}

\put(196,60){$\otimes$}
\put(204,65){$O6^-$}
\put(204,51){$(x+m)$}
\put(220,0){\makebox(0,0)[b]{\bf {Figure 10}}}

\end{picture}
\vspace{.2in}

Therefore, to begin with, for ${\rm SU}(N)$ gauge theory with two 
antisymmetric representations
of masses $m_1$ and $m_2$, 
we expect {\it at least} the brane structure in
Fig.~11.

\begin{picture}(400,150)(10,10)

\put(80,20){\line(0,1){100}}
\put(170,20){\line(0,1){100}}
\put(250,20){\line(0,1){100}}
\put(330,20){\line(0,1){100}}

\put(166,60){$\otimes$}
\put(174,65){$O6^-$}
\put(174,51){$(x+m_2)$}

\put(246,60){$\otimes$}
\put(254,65){$O6^-$}
\put(254,51){$(x+m_1)$}
\put(220,0){\makebox(0,0)[b]{\bf {Figure 11}}}
\end{picture}

\vspace{.2in}

Again in  Fig.~11, only the \hbox{NS 5-branes} and $O6^-$
are shown, while the $N$ \hbox{D4-branes} connecting the \hbox{NS 5-branes} 
and the flavor \hbox{D6-branes}  are not shown for clarity. The
first observation is that to satisfy all possible mirrors,
one must have an infinite chain of \hbox{NS 5-branes} 
and $O6^-$ orientifolds, since one must satisfy the reflections
in {\it each} of the $O6^{-}$ orientifold planes
separately. A portion of this chain is shown in Fig.~12,
which differs from Fig.~11 in that the positions 
of \hbox{D4-branes} and D6-(flavor) branes are shown. 
Fig.~13 is an expanded version of the chain showing 
six \hbox{NS 5-branes}.
One can check that all the necessary 
mirrors about any given $O6^-$
orientifold plane are satisfied. 

There are two important observations:

\noindent
1) If $m_2 \rightarrow \infty$, most of the \hbox{D4-branes},
\hbox{D6-branes} and $O6^-$ planes slide off to infinity, leaving 
us with the configuration of Fig.~14, which  coincides
with Fig.~10 for ${\rm SU}(N)$ and an antisymmetric representation of mass
$m_1$. This is the geometric analogue of the double
scaling limit by which we checked ${\cal F}_{\rm 1-inst}$
for $S(x)$ and (\ref{eifour}).
(Of course, we equally could have taken  $m_1 \rightarrow \infty$.)

2) The ratio  (\ref{hfive})  is characteristic of a single cell of 
two adjacent \hbox{NS 5-branes}, and $N$ linking \hbox{D4-branes}. The
statement that the ratio (\ref{hfive}) is ${1\over 4} S(x)$ up to a
known reflection and shift in $x$ is equivalent to saying that
one can begin the hyperelliptic approximation
with {\it any} cell in the infinite chain of
\hbox{NS 5-branes}.

Thus the infinite chain of \hbox{NS 5-branes} and $O6^-$ orientifolds,
a portion of which is shown in Fig.~13, is equivalent to the
curve (\ref{eiseven}), with solution (\ref{hfive})-({\ref{hseven}), etc. 
The question arises whether
the infinite series can be summed. For simplicity
we consider ${\rm SU}(N)$ with two antisymmetric representations
and $N_f=0$, with $m_1=m_2=m$. The series (\ref{eiseven}), with coefficients
$P_n(x)$ known to {\it leading}  order in $L$,
and $j_n(x)=0$, can be summed. These leading
order terms sum to 
\be
H_0(x)\,\sum_{n=-\infty}^{\infty} e^{2 \pi i \tau(x) n^2} t^{2n}\,+\,
H_1(x)\,\sum_{n=-\infty}^{\infty} e^{2 \pi i \tau(x) (n+1/2)^2} t^{2n+1}\,=\,0,
\label{height}
\ee
where
\be
H_0(x)&=&\prod_{i=1}^N (x-a_i-m),\nn\\
H_1(x)&=&H_0(-x),\nn\\
e^{2 \pi i \tau(x)}\,&=&\,{L^4\over x^8}. 
\label{hnine}
\ee
where the definition of $x$ has been shifted by $m$ 
relative to eq. (\ref{eiseven}).
Eq. (\ref{height}) can be reexpressed in terms of Jacobi theta
functions as 
\be
H_0(x)\,\theta_3(2 \nu|2 \tau(x))\,+\,
H_1(x)\,\theta_2(2 \nu|2 \tau(x))\,=\,0,
\label{hhten}
\ee
where $t=e^{2 \pi i \nu}$,
and \cite{Polchinski}  
\be
\theta_1(\nu|\tau) &=& i \sum_{n=-\infty}^{\infty} (-1)^n 
e^{i \pi \tau (n - \tshalf)^2} e^{2 \pi i \nu (n - \tshalf)}, \nn\\ 
\theta_2(\nu|\tau) &=&  \sum_{n=-\infty}^{\infty} 
e^{i \pi \tau (n - \tshalf)^2} e^{2 \pi i \nu (n - \tshalf)}, \nn\\ 
\theta_3(\nu|\tau) &=&  \sum_{n=-\infty}^{\infty} 
e^{i \pi \tau n^2} e^{2 \pi i \nu n}, \nn\\ 
\theta_4(\nu|\tau) &=&  \sum_{n=-\infty}^{\infty} (-1)^n 
e^{i \pi \tau n^2} e^{2 \pi i \nu n} .
\label{thetafcns}
\ee
Note that (\ref{hhten}) only includes the leading terms of the curve.
Since Im $\tau(x) > 0 $ is required
for the theta functions, the series is not well defined
for $L$ large or $x \rightarrow 0$, the latter describing 
the approach to an orientifold. 
This clearly indicates the need for subleading terms 
({\it i.e.} terms of higher order in $e^{2\pi i \tau(x)}$ 
for a given power of $t$) 
in order to ``resolve" the singularities at the orientifold. 
These are non-perturbative issues. 
Also note that (\ref{hhten}) is reminiscent of, 
but not identical to, the spectral 
curve of the Calogero-Moser model, which describes
${\rm SU}(N)$ with an adjoint matter hypermultiplet 
\cite{MartinecWarner, DHokerPhong}. 

Another case for which the infinite series (\ref{eiseven}) can 
be summed is that for which we extrapolate
to $N_f=4$, and take $m_1=m_2=M_j,\, \,\,(j=1,...,4)$. This
is an {\it elliptic} model, of which more will be 
discussed in the next section. Using our solution
for the coefficients $P_n(x)$ and $j_n(x)$ we obtain
\be
H_0(x)\,\theta_3(2 \nu|2 \tau)\,+\,
H_1(x)\,\theta_2(2 \nu|2 \tau)\,=\,0,
\label{hten}
\ee
where now $\tau$ is  independent of $x$, and
is the modular parameter of the elliptic model. 
Thus, we are dealing with 
a mass-deformed  scale-invariant case, {\it i.e.} a theory with zero
beta function.  We return to this case
in the next section, where we indicate in (\ref{htfour})--(\ref{htnine})
that there are no subleading terms in (\ref{hten}).

\begin{center}
\begin{picture}(810,295)(10,10)


\put(10,192){\line(1,0){5}}
\put(20,192){\line(1,0){5}}
\put(30,192){\line(1,0){5}}
\put(40,192){\line(1,0){5}}
\put(50,192){\line(1,0){5}}
\put(60,192){\line(1,0){5}}
\put(70,192){\line(1,0){5}}
\put(80,192){\line(1,0){5}}
\put(90,192){\line(1,0){5}}
\put(100,192){\line(1,0){5}}
\put(5,180){$(\!x\!-\!a_i\!+\!2m_2\!-\!2m_1\!)$}

\put(105,240){\line(1,0){2}}
\put(110,240){\line(1,0){5}}
\put(120,240){\line(1,0){5}}
\put(130,240){\line(1,0){5}}
\put(140,240){\line(1,0){5}}
\put(150,240){\line(1,0){5}}
\put(160,240){\line(1,0){5}}
\put(170,240){\line(1,0){5}}
\put(180,240){\line(1,0){5}}
\put(190,240){\line(1,0){5}}
\put(200,240){\line(1,0){5}}
\put(120,245){$(\!x\!+\!a_i\!+\!2m_2\!)$}

\put(205,117){\line(1,0){2}}
\put(210,117){\line(1,0){5}}
\put(220,117){\line(1,0){5}}
\put(230,117){\line(1,0){5}}
\put(240,117){\line(1,0){5}}
\put(250,117){\line(1,0){5}}
\put(260,117){\line(1,0){5}}
\put(270,117){\line(1,0){5}}
\put(280,117){\line(1,0){5}}
\put(290,117){\line(1,0){5}}
\put(300,117){\line(1,0){5}}
\put(235,105){$(\!x-\!a_i\!)$}

\put(305,155){\line(1,0){2}}
\put(310,155){\line(1,0){5}}
\put(320,155){\line(1,0){5}}
\put(330,155){\line(1,0){5}}
\put(340,155){\line(1,0){5}}
\put(350,155){\line(1,0){5}}
\put(360,155){\line(1,0){5}}
\put(370,155){\line(1,0){5}}
\put(380,155){\line(1,0){5}}
\put(390,155){\line(1,0){5}}
\put(400,155){\line(1,0){5}}
\put(320,160){$(\!x\!+\!a_i\!+\!2m_1\!)$}

\put(405,10){\line(1,0){2}}
\put(410,10){\line(1,0){5}}
\put(420,10){\line(1,0){5}}
\put(430,10){\line(1,0){5}}
\put(440,10){\line(1,0){5}}
\put(450,10){\line(1,0){5}}
\put(460,10){\line(1,0){5}}
\put(470,10){\line(1,0){5}}
\put(480,10){\line(1,0){5}}
\put(490,10){\line(1,0){5}}
\put(500,10){\line(1,0){5}}
\put(407,15){$(\!x\!-\!a_i\!+\!2m_1\!-\!2m_2\!)$}


\put(105,5){\line(0,1){255}}
\put(205,5){\line(0,1){255}}
\put(305,5){\line(0,1){255}}
\put(405,5){\line(0,1){255}}
\put(103,271){$\NSone$}
\put(203,271){$\NStwo$}
\put(303,271){$\NSthree$}
\put(403,271){$\NSfour$}


\put(101,215){$\otimes$}
\put(201,175){$\otimes$}
\put(301,135){$\otimes$}
\put(401,85){$\otimes$}
\put(31,220){$(\!x\!+\!2m_2\!-\!m_1\!)$}
\put(162,170){$(\!x\!+\!m_2\!)$}
\put(262,135){$(\!x\!+\!m_1\!)$}
\put(406,75){$(\!x\!+\!2m_1\!-\!m_2\!)$}

\put(151,185){\framebox(5,5){$\cdot$}}
\put(251,165){\framebox(5,5){$\cdot$}}
\put(351,105){\framebox(5,5){$\cdot$}}
\put(120,192){$(\!x\!+\!2m_2\!-\!M_j\!)$}
\put(235,172){$(\!x\!+\!M_j\!)$}
\put(320,96){$(\!x+\!2m_1\!-\!M_j\!)$}

\put(30,30){\vector(1,0){20}}
\put(30,30){\vector(0,1){20}}
\put(25,50){$x$}
\put(50,25){$t$}

\end{picture}
\end{center}
\label{figureone}
{\footnotesize{\bf Figure 12}: \hfil\break
\noindent 1) {\it vertical lines}: 
parallel, equally spaced \hbox{NS 5-branes}.

\noindent
2) {\it dashed lines}: 
$N$ parallel \hbox{D4-branes} connect pairs of adjacent \hbox{NS 5-branes}.

\noindent
3) $\otimes$: $O6^{-}$ orientifold planes.

\noindent
4) ${\framebox(5,5){$\cdot$}}\,$:  D6-(flavor) branes.

\noindent
Due to mirrors, the picture must extend infinitely to right and left.}
\vfil
\eject

\begin{center}
\begin{picture}(810,400)(10,10)


\put(10,395){\line(1,0){5}}
\put(20,395){\line(1,0){5}}
\put(30,395){\line(1,0){5}}
\put(40,395){\line(1,0){5}}
\put(50,395){\line(1,0){5}}
\put(60,395){\line(1,0){5}}
\put(5,400){$\text{\scriptstyle{(\!x\!+\!a_i\!+\!4m_2-\!2m_1\!)}}$}

\put(65,240){\line(1,0){2}}
\put(70,240){\line(1,0){5}}
\put(80,240){\line(1,0){5}}
\put(90,240){\line(1,0){5}}
\put(100,240){\line(1,0){5}}
\put(110,240){\line(1,0){5}}
\put(120,240){\line(1,0){5}}
\put(130,240){\line(1,0){5}}
\put(72,245){$\text{\scriptstyle{(\!x\!-\!a_i\!+\!2m_2\!-\!2m_1\!)}}$}

\put(135,295){\line(1,0){2}}
\put(140,295){\line(1,0){5}}
\put(150,295){\line(1,0){5}}
\put(160,295){\line(1,0){5}}
\put(170,295){\line(1,0){5}}
\put(180,295){\line(1,0){5}}
\put(190,295){\line(1,0){5}}
\put(200,295){\line(1,0){5}}
\put(150,300){$\text{\scriptstyle{(\!x\!+\!a_i\!+\!2m_2\!)}}$}

\put(205,145){\line(1,0){2}}
\put(210,145){\line(1,0){5}}
\put(220,145){\line(1,0){5}}
\put(230,145){\line(1,0){5}}
\put(240,145){\line(1,0){5}}
\put(250,145){\line(1,0){5}}
\put(260,145){\line(1,0){5}}
\put(270,145){\line(1,0){5}}
\put(228,148){$\text{\scriptstyle{(\!x-\!a_i\!)}}$}

\put(275,190){\line(1,0){2}}
\put(280,190){\line(1,0){5}}
\put(290,190){\line(1,0){5}}
\put(300,190){\line(1,0){5}}
\put(310,190){\line(1,0){5}}
\put(320,190){\line(1,0){5}}
\put(330,190){\line(1,0){5}}
\put(340,190){\line(1,0){5}}
\put(290,193){$\text{\scriptstyle{(\!x\!+\!a_i\!+\!2m_1\!)}}$}

\put(345,45){\line(1,0){2}}
\put(350,45){\line(1,0){5}}
\put(360,45){\line(1,0){5}}
\put(370,45){\line(1,0){5}}
\put(380,45){\line(1,0){5}}
\put(390,45){\line(1,0){5}}
\put(400,45){\line(1,0){5}}
\put(410,45){\line(1,0){5}}
\put(350,48){$\text{\scriptstyle{(\!x\!-\!a_i\!+\!2m_1\!-\!2m_2\!)}}$}

\put(415,90){\line(1,0){2}}
\put(420,90){\line(1,0){5}}
\put(430,90){\line(1,0){5}}
\put(440,90){\line(1,0){5}}
\put(450,90){\line(1,0){5}}
\put(460,90){\line(1,0){5}}
\put(470,90){\line(1,0){5}}
\put(480,90){\line(1,0){5}}
\put(420,93){$\text{\scriptstyle{(\!x\!+\!a_i\!+\!4m_1\!-\!2m_2\!)}}$}


\put(65,5){\line(0,1){405}}
\put(135,5){\line(0,1){405}}
\put(205,5){\line(0,1){405}}
\put(275,5){\line(0,1){405}}
\put(345,5){\line(0,1){405}}
\put(415,5){\line(0,1){405}}
\put(63,421){$\NSzero$}
\put(133,421){$\NSone$}
\put(203,421){$\NStwo$}
\put(273,421){$\NSthree$}
\put(343,421){$\NSfour$}
\put(413,421){$\NSfive$}

\put(26,345){\framebox(5,5){$\cdot$}}
\put(96,285){\framebox(5,5){$\cdot$}}
\put(166,245){\framebox(5,5){$\cdot$}}
\put(236,185){\framebox(5,5){$\cdot$}}
\put(306,145){\framebox(5,5){$\cdot$}}
\put(376,85){\framebox(5,5){$\cdot$}}
\put(446,45){\framebox(5,5){$\cdot$}}
\put(1,337){$\text{\scriptstyle{(\!x\!+\!4m_2\!-\!2m_1\!-\!M_j\!)}}$}
\put(69,277){$\text{\scriptstyle{(\!x\!+\!2m_2\!-\!2m_1\!+\!M_j\!)}}$}
\put(149,237){$\text{\scriptstyle{(\!x\!+\!2m_2\!-\!M_j\!)}}$}
\put(230,177){$\text{\scriptstyle{(\!x\!+\!M_j\!)}}$}
\put(289,137){$\text{\scriptstyle{(\!x+\!2m_1\!-\!M_j\!)}}$}
\put(347,77){$\text{\scriptstyle{(\!x\!+\!2m_1\!-\!2m_2\!+\!M_j\!)}}$}
\put(420,37){$\text{\scriptstyle{(\!x\!+\!4m_1\!-\!2m_2\!-\!M_j\!)}}$}


\put(61,315){$\otimes$}
\put(131,265){$\otimes$}
\put(201,215){$\otimes$}
\put(271,165){$\otimes$}
\put(341,115){$\otimes$}
\put(411,65){$\otimes$}
\put(71,315){$\text{\scriptstyle{(\!x\!+\!3m_2\!-\!2m_1\!)}}$}
\put(141,265){$\text{\scriptstyle{(\!x\!+\!2m_2\!-\!m_1\!)}}$}
\put(211,215){$\text{\scriptstyle{(\!x\!+\!m_2\!)}}$}
\put(281,165){$\text{\scriptstyle{(\!x\!+\!m_1\!)}}$}
\put(351,115){$\text{\scriptstyle{(\!x\!+\!2m_1\!-\!m_2\!)}}$}
\put(421,65){$\text{\scriptstyle{(\!x\!+\!3m_1\!-\!2m_2\!)}}$}

\put(30,30){\vector(1,0){20}}
\put(30,30){\vector(0,1){20}}
\put(25,50){$x$}
\put(50,25){$t$}


\end{picture}
\end{center}
\label{figuretwo}
{\footnotesize{\bf Figure 13}: 
An expanded version of Figure 12 with six \hbox{NS 5-branes}. 
One can do hyperelliptic 
approximation about {\it any} pair of adjacent \hbox{NS 5-branes}.}

\eject

\begin{center}
\begin{picture}(810,295)(10,10)


\put(195,117){\line(1,0){2}}
\put(200,117){\line(1,0){5}}
\put(210,117){\line(1,0){5}}
\put(220,117){\line(1,0){5}}
\put(230,117){\line(1,0){5}}
\put(240,117){\line(1,0){5}}
\put(250,117){\line(1,0){5}}
\put(260,117){\line(1,0){5}}
\put(268,117){\line(1,0){2}}
\put(210,105){$(\!x-\!a_i\!)$}

\put(270,155){\line(1,0){2}}
\put(275,155){\line(1,0){5}}
\put(285,155){\line(1,0){5}}
\put(295,155){\line(1,0){5}}
\put(305,155){\line(1,0){5}}
\put(315,155){\line(1,0){5}}
\put(325,155){\line(1,0){5}}
\put(335,155){\line(1,0){5}}
\put(343,155){\line(1,0){2}}
\put(280,160){$(\!x\!+\!a_i\!+\!2m_1\!)$}


\put(45,5){\line(0,1){235}}
\put(120,5){\line(0,1){235}}
\put(195,5){\line(0,1){235}}
\put(270,5){\line(0,1){235}}
\put(345,5){\line(0,1){235}}
\put(420,5){\line(0,1){235}}
\put(43,251) {$\NSzero$}
\put(118,251) {$\NSone$}
\put(193,251){$\NStwo$}
\put(268,251){$\NSthree$}
\put(343,251) {$\NSfour$}
\put(418,251) {$\NSfive$}


\put(266,135){$\otimes$}

\put(227,135){$(\!x\!+\!m_1\!)$}


\put(231,165){\framebox(5,5){$\cdot$}}
\put(306,105){\framebox(5,5){$\cdot$}}

\put(215,172){$(\!x\!+\!M_j\!)$}
\put(272,96){$(\!x+\!2m_1\!-\!M_j\!)$}

\put(10,30){\vector(1,0){20}}
\put(10,30){\vector(0,1){20}}
\put(05,50){$x$}
\put(30,25){$t$}

\end{picture}
\end{center}
\label{figurethree}
{\footnotesize{\bf Figure 14}: The $m_2\to \infty$ limit of Figure 13. 
In this limit, only the \hbox{NS 5-branes} $\NStwo$, $\NSthree$, and  $\NSfour$ 
remain connected by \hbox{D4-branes}. 
The other \hbox{D4-branes} and $O6^{-}$ planes have ``slid off'' to $x \sim \infty$.
It agrees with the \hbox{M-theory} picture of 1 antisymmetric hypermultiplet 
with mass $2m_1$.}

\eject

\noindent{\bf 8. ~Elliptic Models}
\renewcommand{\theequation}{8.\arabic{equation}}
\setcounter{equation}{0}

We remarked that (\ref{hten}) was reminiscent of the Calogero-Moser model. 
Let us consider this in more detail, by examining
${\cal N}=2$ ${\rm SU}(N)$ gauge theory 
with an adjoint hypermultiplet of mass $m$. 
This scale invariant model can be expressed 
in terms of an \hbox{M-theory} picture \cite{Witten}:

\begin{figure} [h]
\centerline{
\psfig{figure=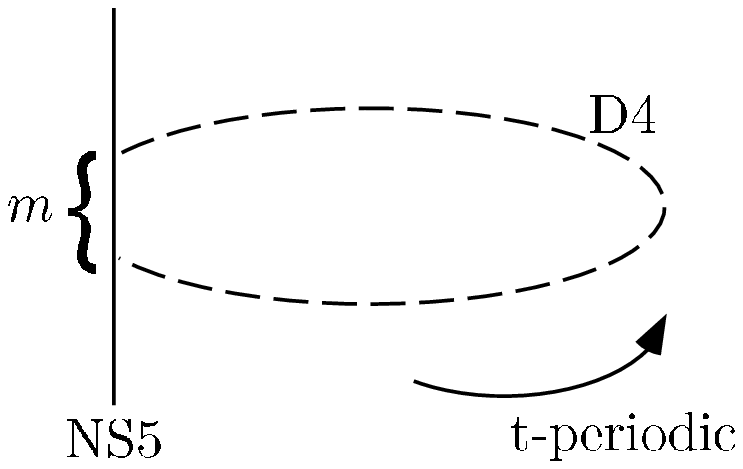,height=3.6cm,width=4.6cm}}
\begin{center}
{\footnotesize{\bf Figure 15}}
\end{center}
\end{figure}

In Fig.~15, there are $N$ \hbox{D4-branes} suspended between 
a single \hbox{NS 5-brane}, with a periodicity in $t$, but with a shift
in $v$ an ammount $m$ for each circuit of $t$. Thus, there is a global mass
$m$. The covering space of the $S^1$ (the $t$-variable) is shown in Fig.~16.

\begin{picture}(600,140)(10,10)

\put(40,40){\line(0,1){100}}
\put(130,40){\line(0,1){100}}
\put(220,40){\line(0,1){100}}
\put(310,40){\line(0,1){100}}
\put(400,40){\line(0,1){100}}

\put(-10,50){\line(1,0){10}}
\put(10,50){\line(1,0){10}}
\put(30,50){\line(1,0){10}}

\put(40,65){\line(1,0){10}}
\put(60,65){\line(1,0){10}}
\put(80,65){\line(1,0){10}}
\put(100,65){\line(1,0){10}}
\put(120,65){\line(1,0){10}}

\put(130,80){\line(1,0){10}}
\put(150,80){\line(1,0){10}}
\put(170,80){\line(1,0){10}}
\put(190,80){\line(1,0){10}}
\put(210,80){\line(1,0){10}}

\put(220,95){\line(1,0){10}}
\put(240,95){\line(1,0){10}}
\put(260,95){\line(1,0){10}}
\put(280,95){\line(1,0){10}}
\put(300,95){\line(1,0){10}}

\put(310,110){\line(1,0){10}}
\put(330,110){\line(1,0){10}}
\put(350,110){\line(1,0){10}}
\put(370,110){\line(1,0){10}}
\put(390,110){\line(1,0){10}}

\put(400,125){\line(1,0){10}}
\put(420,125){\line(1,0){10}}
\put(440,125){\line(1,0){10}}

\put(100,25){\vector(1,0){35}}
\put(140,20){${\it t\ (\rm{covering\ of}\ S^1)}$}

\put(10,100){\vector(0,1){30}}
\put(15,110){$v$}
\put(220,0){\makebox(0,0)[b]{\bf {Figure 16}}}

\end{picture}
\vspace{.2in}

Generalizing the rules of Table 1, we expect from Fig.~16 
the function 
\be
S(v)\,=\,{{\prod_{i=1}^N (v-a_i-m)\,\prod_{i=1}^N (v-a_i+m)} 
\over {\prod_{i=1}^N (v-a_i)^2}},
\label{hfifteen}
\ee
which has the property expressed in (\ref{hfive}), 
but with no subleading terms. 
The instanton expansion of ${\rm SU}(N)$ +
massive adjoint representation is given in \cite{DHokerPhong, 
Minahan}.

Witten \cite{Witten} shows that the curve for this model is 
precisely that derived by Donagi and Witten \cite{DonagiWitten}
in the context of the integrable Hitchin system
\be
F(v,x,y)\,=\,\sum_{j=0}^N A_j P_{N-j}(v),
\label{hsixteen}
\ee
where $A_j$ are 
gauge invariant polynomials in the  $v$
(the vev  of the scalar field), 
and where $x$ and $y$ are related by the elliptic curve
\be
y^2\,=\,(x-e_1)(x-e_2)(x-e_3).
\label{hseventeen}
\ee
They show that 
\be
P_n(v)\,=\,\sum_{i=0}^n\,
\left( n \atop i \right) f_i \,\, v^{n-i}.
\label{heighteen}
\ee
It can be shown that (\ref{hsixteen}) can be reexpressed as 
\be
F(v,x,y)\,=\,\sum_{j=0}^N {m^j \over j!}\,\, f_j\,\, H^{(j)}(t),
\label{hnineteen}
\ee
where
\be
H(v)=\prod_{i=1}^N (v-a_i)\,\,,\,\,\,\,\,\,\,\,\,\,
H^{(j)}(v)= {d^j H(v) \over dv^j}.
\label{htwenty}
\ee
Explicit calculation gives
\be
f_0&=&1\,\,\,,\,\,\,\,\,\,\,\,\,\,\,
f_1\,=\,0\,\,\,,\,\,\,\,\,\,\,\,\,\,
f_2\,=\,-x\,,\,\,\,\,\,\,\,\nn\\
f_3&=&2y\,\,\,,\,\,\,\,\,\,\,\,\,\,
f_4\,=\,-3x^2+4x\sum_{i=1}^3 e_i\,,\,\,\,\,\,\,\,\,\,\, {\rm etc}.
\label{htone}
\ee

It is claimed that this is equivalent to the Calogero-Moser
model in ref. \cite{DHokerPhong},
for which the spectral curve is
\be
\sum_{n=-\infty}^{\infty}\,(-1)^n q^{{1\over 2} n(n-1)} e^{nz} H(k-nm)\,=\,0, 
\label{htwelve}
\ee
where $z$ parametrizes the torus with 
the identifications 
$z = z-2\pi i = z - 2\pi i\tau$
and $q=e^{2\pi i \tau}$.
Heretofore
the connection between these curves has not been made explicit. 
We will sketch our results in this regard \cite{ELNS}. 
The curve (\ref{htwelve}) can be recast as 
\be
\sum_{j=0}^{N}\,{(-m)^j \over j!}\,h_j(z)  H^{(j)} (k-{\tshalf}m)\,=\,0, 
\label{hthirteen}
\ee 
where
\be
h_j(z)={1 \over {\theta_1({z \over -2 \pi i}|\tau)}}{\pr^j \over {\pr z^j}}
\,\theta_1({z \over -2 \pi i}|\tau), 
\label{hfourteen}
\ee 
with $\theta_1$ defined in eq.~(\ref{thetafcns}).
Upon a change of variables,
equation (\ref{hthirteen}) becomes
\be
\sum_{j=0}^N {m^j \over j!} \, \, \tf_j(z) \, \,H^{(j)}(v) = \, 0.
\ee
The functions $f_j$ and $\tf_j$ can be shown to be equivalent
by using the connection
\be
x\,=\,{\wp}(z)\,\,,\,\,\,\,\,\,\,\,\,\, 2y\,=\, {\wp}'(z),
\label{nttwo}
\ee
where ${\wp}(z)$ is the Weierstrass elliptic function.

Another important elliptic model is that of ${\rm SU}(2N)$ 
with two antisymmetric representations and $N_f=4$, 
with masses satisfying $m_1=m_2=M_j=m$, as in (\ref{hten}).
This is an untwisted elliptic model with zero global mass. 
\begin{figure} [h]
\centerline{ \psfig{figure=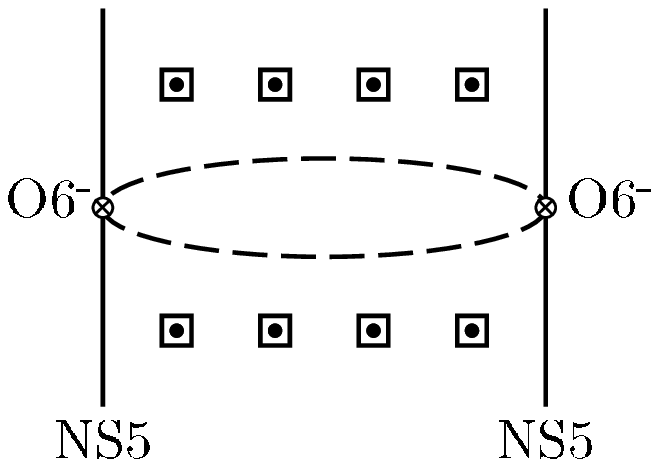,height=3.6cm,width=5cm}}
\begin{center}
{\footnotesize{\bf Figure 17}}
\end{center}
\end{figure}
The brane picture has been discussed by Uranga \cite{Uranga},
and is shown in Fig.~17.
The curve appropriate to Fig.~17
has been given by Gukov and Kapustin \cite{GukovKapustin} as
\be
v^{2N}\,+\,f_1(x,y) v^{2N-1}\,+ \,\cdots\, +\, f_{2N}(x,y)
\,=\,0,
\label{htfour}
\ee
The coefficient functions are given by
\be
f_{2j}(x,y)&=&A_j,\nn\\
f_{2j-1}(x,y)&=&{{yB_j}\over (x-e_3)}\,=\,{{(x-e_1)(x-e_2)}\over y} B_j ,
\label{htfive}
\ee
where $x$ and $y$ satisfy (\ref{hseventeen}) 
and $A_j$ and $B_j$ are constants. 
We have shown \cite{ELNS} that 
(\ref{htfour}) and (\ref{htfive}) may be recast as 
\be
& &H_+(v) \,{\rm tan} (\pi \nu)  \prod_{n=1}^{\infty}
(1\,-\,2e^{i\pi n \tau} \cos 2\pi\nu\,+\,e^{2\pi i n \tau}) \nn\\
&+& H_-(v) [\theta_4(0|\tau)]^2\,\prod_{n=1}^{\infty}
(1\,+ \,2e^{i\pi n \tau} \cos 2\pi\nu\,+\,e^{2\pi i n \tau}) \,=\,0,\nn\\
\label{htsix}
\ee
with
\be
H_{\pm} (v)&=&{\tshalf}(H_0\,\pm\,H_1),\nn\\
H_0(v)  &=& \prod_{j=1}^{2N}(v-a_j-m),\nn\\
H_1(v)  &=& H_0(-v).
\label{hteight}
\ee
We then found that (\ref{htsix}) is equivalent to
\be
H_0(v)\,\theta_3(2 \nu|2 \tau)\,+\,
H_1(v)\,\theta_2(2 \nu|2 \tau)\,=\,0,
\label{htnine}
\ee
after a suitable change of variables. 
This is {\it identical} to (\ref{hten}).

Thus, beginning with an elliptic model, and only
analytic tools, we have a very dramatic confirmation of 
the reasoning of Sec. 7,
which involved both analytic and geometric reasoning!

There are other interesting elliptic models, but we do not discuss these here.

On general principles, one expects there to be an integrable
model associated to the solution of the \hbox{${\cal N}=2$} 
SW problem (see Gorsky et. al. \cite{Gorsky}
for known examples). For example, 
${\rm SU}(N)$ with adjoint matter
was discussed in the first part of this section, and the explicit
connection between the spectral curve of the Calogero-Moser model and
the Donagi-Witten construction was sketched. However, there are a number of
SW problems for which there are no known integrable models. A typical case in
point is that of ${\rm SU}(N)$ with two antisymmetric representations and
$N_f=4$, with a single common mass $m$, discussed in the second part
of this section. It would be very helpful to make these identifications, both 
on fundamental grounds, and also because the techniques of 
the integrable models may be able to provide
new  non-perturbative results, which go beyond 
hyperelliptic perturbation theory,
carrying us into other regions of moduli space.

\noindent{\bf 9. ~Concluding Remarks}
\renewcommand{\theequation}{9.\arabic{equation}}
\setcounter{equation}{0}

There are a number of open problems which should be addressed. An incomplete 
list is: 

\noindent
1) Compute ${\cal F}_{\rm 1-inst}$ from ${\cal L}_{\rm micro}$ for all 
the cases described in Table 1, so as to extend the test of \hbox{M-theory}.
In every case where a test can be made, agreement has been found. 

\noindent
2) Explain {\it group-theoretically} the entries
for $S(x)$ in Table 1, and the rules (\ref{eighty})-(\ref{eithree}) 
abstracted from this table.

\noindent
3) Study other elliptic models along the lines described in this review. 

\noindent
4) Find the associated integrable models for all the cases 
describable by a SW curve.

\noindent
5) Extend the predictions of non-hyperelliptic curves 
to regions of moduli space for which the hyperelliptic perturbation 
theory is not valid. 

As we have discussed, \hbox{${\cal N}=2$} SW theory presents 
many varied opportunities for testing \hbox{M-theory} predictions for gauge 
theories. These deserve to be explored further to increase 
our confidence in \hbox{M-theory}.
\eject

\noindent{\bf 10. ~Acknowledgement}
\renewcommand{\theequation}{10.\arabic{equation}}
\setcounter{equation}{0}

H.J.S. wishes to thank the organizers J.M.F. Labastida
and J. Barb\'on of the {\it Advanced School of Supersymmetry in the 
Theories of Fields,  Strings and Branes} for the opportunity
to present the results of our group in such a stimulating atmosphere,
and beautiful and exciting environment. This review represents joint work of
a very fruitful collaboration. 

The authors wish to thank Carlos Lozano 
for his collaboration on material in Section 8.

\eject

\end{document}